\documentclass{aa}  

\usepackage{graphicx}
\usepackage{txfonts}
\usepackage[dvipsnames]{xcolor}
\usepackage[figuresright]{rotating}
\usepackage{hyperref}
\hypersetup{
    colorlinks=true,
    allcolors=blue
    }
\usepackage{tikz}
\usetikzlibrary{shapes.geometric, arrows}
\tikzstyle{io} = [rectangle, rounded corners, minimum width=2.5cm, text width=3cm, minimum height=1cm, text centered, draw=black, fill=red!30]
\tikzstyle{function} = [rectangle, minimum width=3cm, text width=3cm, minimum height=1cm, text centered, draw=black, fill=orange!30]
\tikzstyle{subfunction} = [rectangle, rounded corners=0.5cm, minimum width=3cm, text width=3cm, minimum height=1cm, text centered, draw=black, fill=yellow!50]
\tikzstyle{result} = [trapezium, 
    trapezium left angle = 65,
    trapezium right angle = 115,
    trapezium stretches, minimum width=3cm, text width=3cm, minimum height=1cm, text centered, draw=black, fill=green!30]
\tikzstyle{condition} = [diamond, aspect=3, minimum width=2.2cm, text width=2.2cm, minimum height=1cm, text centered, draw=black, fill=blue!30]
\tikzstyle{arrow} = [thick,->,>=stealth]

\usepackage{ulem}

\begin{document}

\title{An innovative and automated method for vortex identification}
\subtitle{I. Description of the SWIRL algorithm}

\author{
      J.~R.~Canivete Cuissa\inst{1,2}
      \and
      O.~Steiner\inst{1,3}
}

\institute{
        IRSOL Istituto Ricerche Solari ``Aldo e Cele Daccò'' Locarno, Università della Svizzera Italiana (USI), \\
        Via Patocchi 57 -- Prato Pernice, 6605 Locarno-Monti, Switzerland\\
        \email{jose.canivete@irsol.usi.ch}
    \and        
        Center for Theoretical Astrophysics and Cosmology, Institute for Computational Science (ICS),\\University of Zurich, Winterthurerstrasse 190, 8057 Z{\"u}rich, Switzerland
     \and 
        Leibniz-Institut f\"ur Sonnenphysik (KIS),\\ 
        Sch\"oneckstrasse 6, 79104 Freiburg i.Br., Germany
}

\date{Received 11 April 2022 / Accepted 07 October 2022}

\abstract
{
A universally accepted definition of what a vortex is has not yet been reached. Therefore, we lack an unambiguous and rigorous method for the identification of vortices in fluid flows.
Such a method would be necessary to conduct robust statistical studies on vortices in highly dynamical and turbulent systems, such as the solar atmosphere.
}
{
We aim to develop an innovative and robust automated 
methodology for the identification of vortices based on local and global characteristics of the flow.
Moreover, the use of a threshold 
that could potentially prevent the detection of weak vortices in the identification process should be avoided.
}
{
We present a new method that combines the rigor of mathematical criteria with the global perspective of morphological techniques. The core of the method consists in the estimation of the center of rotation for every point of the flow that presents some degree of curvature in its neighborhood. For that, we employ the Rortex criterion and 
combine it with morphological considerations of the velocity field.
We then identify coherent vortical structures by clusters of estimated centers of rotation.
}
{
We demonstrate that the Rortex is a more reliable criterion than are the swirling strength and the vorticity
for the extraction of physical information from vortical flows, because it measures the rigid-body rotational part of the flow alone and is not biased by the presence of pure or intrinsic shears. We show that the method performs well on a simplistic test case composed of two Lamb-Oseen vortices.
We combine the proposed method with a state of the art clustering algorithm to build an automated vortex identification algorithm. 
The algorithm is applied to an artificial flow composed of multiple Lamb-Oseen vortices with a random noisy background and to the turbulent flow of a simulated magneto-hydrodynamical Orszag-Tang vortex test.
The results demonstrate the reliability and accuracy of the method. A Python implementation of the algorithm is publicly available.
}
{
The present automated vortex identification method can be considered a new tool for the detection and study of vortices in dynamical and turbulent (magneto-)hydrodynamical flows. By applying the implemented algorithm to numerical simulations and observational data, and by comparing it to existing detection methods, we seek to successively improve the reliability of the detections and, ultimately, our knowledge on swirling motions in the solar, stellar, and planetary atmospheres.
}


\titlerunning{An innovative and automated vortex identification method I}

\maketitle
   
%
%
\section{Introduction}
\label{sec:introduction}

Vortices are one of the fundamental features of fluid dynamics and turbulent flows. Although a vortex can conceptually be described as a fluid region rotating around a common axis, the task of mathematically defining what a vortex is has proven to be incredibly challenging and it is still subject of debate \citep[see, e.g.,][]{2018CGF...37...6G}. 
A rigorous and objective definition is particularly needed for the identification of vortices in highly dynamical and turbulent flows, where detections based on the naive human intuition of swirling structures could bias the results.

Multiple detection methods and criteria have been proposed in the literature. 
Most of them can be categorized into two classes. 
The first and most widely known class consists on mathematical criteria based on local physical quantities related to the flow, such as the velocity field, the pressure, and their derivatives.
Local criteria are defined for each point in space. In practice, on a discrete grid, the vortex identification involves a limited stencil of neighboring grid cells.
Examples are
the vorticity $\vec{\omega}$, 
the $Q$-criterion \citep{1988stun.proc..193H}, 
the $\lambda_2$-criterion \citep{1995JFM...285...69J}, 
the swirling strength $\lambda$ \citep{1999JFM...387..353Z}, 
the $\Gamma$ functions \citep{2001MeScT..12.1422G}, 
and the instantaneous vorticity deviation/Lagrangian-averaged vorticity deviation (IVD/LAVD) \citep{2016JFM...795..136H}. 
A vortex is usually identified as a connected over-density region of one of these criteria.

Most of these methods are mathematically rigorous and physically consistent. For example, all the listed criteria are Galilean invariant and the IVD/LAVD is considered an objective measure\footnote{An objective measure is invariant under changes of reference frame. Mathematically, it translates into invariance with respect to all Euclidean transformations \citep{2016JFM...795..136H}.} \citep{2016JFM...795..136H}. 
However, part of these methods may be prone to miss vortices that are characterized by weak rotational velocities, since mathematical criteria are related to the angular velocity of the flow and a threshold is usually employed to filter out low-magnitude, noisy signals.
This problem concerns in particular the vorticity, the $Q$-criterion, the $\lambda_2$-criterion, the swirling strength, for which weak rotational signals and the background, turbulent noise are essentially indistinguishable, while the $\Gamma$ functions method by definition requires a threshold on the value of $|\Gamma_1|$. The IVD/LAVD methods may be more robust in this regard as they are based on fluctuations of the vorticity with respect to the domain spatial mean, which should include to a certain degree the background noise.  
Moreover, false detections can happen when the flow is curved but does not perform a full rotation, since these methods only measure the local curvature at each grid point and do not take into account the global behavior of the flow. 

The second class consists of morphological methods, which take properties of the flow in the entire domain (or a subset of it) into account. Morphological criteria are not necessarily defined in each point of space as vortices may be sparsely distributed.
These methods rely on intuitive definitions of what a vortex is, such as ``a vortex consists in a multitude of material particles rotating around a common center'' \citep{1979rdte.book..309L}, 
or ``a vortex exists when instantaneous streamlines mapped onto a plane normal to the vortex core exhibit a roughly circular or spiral pattern, when viewed from a reference frame moving with the
center of the vortex core'' \citep{Robinson1990}. 

\citet{Sadar99} 
presented two techniques for the identification of vortices based on characterization of streamlines. The first one, the ``curvature center method'', determines the center of curvature of streamlines and defines a vortex core as a region where curvature centers accumulate. 
The second, the ``winding-angle method'', detects vortices by clustering  streamlines that appear to be curved around a close set of points. 
These methods are conceptually simple and consider the global features of the flow, but they usually are computationally expensive and are not easily extendable to three dimensions. 
Moreover, streamlines can give false impressions of vortical motions in unsteady and highly dynamical flows  \citep[see, e.g.,][]{2013ApJ...776L...4S}. 

Many of the listed methods have been employed to identify vortices in observations and numerical simulations of the solar atmosphere. Small-scale vortical motions appear to be ubiquitous in the quiet solar photosphere and chromosphere. Moreover, they are thought to play an important role in the transport of energy towards the upper layers of the solar atmosphere \citep[see, e.g,][for a review]{2022...ISSI}. 

Numerical simulations of the solar atmosphere are particularly suited for the application of mathematical criteria for the identification of vortices, since the velocity field and other quantities of interest are directly accessible at each point in the three-dimensional space. 
The vorticity was used by \citet[][]{1998ApJ...499..914S}, \citet[][]{2011A&A...526A...5S}, \citet{2013ApJ...776L...4S}, and \citet[][]{2012ASPC..456....3S}, 
to detect vortices and study their dynamics, while the enstrophy, defined as $|\boldsymbol{\omega}|^2$, was used by \citet[][]{2012ApJ...751L..21K} 
to visualize three-dimensional vortical structures. 
\citet[][]{2011A&A...533A.126M} 
firstly introduced the swirling strength in the context of photospheric swirls, quantity adopted also by  \citet[][]{2012A&A...541A..68M}, \citet[][]{2017A&A...601A.135K}, \citet[][]{2020A&A...639A.118C}, \citet[][]{2020ApJ...894L..17Y}, and \citet[][]{2021A&A...649A.121B}. 
More recently, also the IVD and LAVD criteria have been employed in numerical simulations of the solar photosphere and chromosphere by \citet[][]{2020ApJ...898..137S}, \citet{2021ApJ...915...24S}, 
and \citet[][]{2022ApJ...928....3A}. 

Vortex identification methods can not be straightforwardly applied to observations since the necessary quantities are not immediately available from the data. However, it is possible to estimate the horizontal velocity field through the use of local correlation tracking (LCT) techniques. Therefore, identification methods that rely on the velocity field alone have also been employed on observational data. 
From horizontal velocity maps derived with LCT methods, \citet[][]{2017ApJS..229...14R} and \citet{2018A&A...610A..84R} 
employed Lagrangian tracers to visually identify vortices in super and meso-granular vertices, while 
\citet[][]{2018ApJ...869..169G}, \citet[][]{2019ApJ...872...22L}, and \citet{2019NatCo..10.3504L} 
utilized the $\Gamma$ functions. \citet[][]{2018ApJ...863L...2S} 
presented a comparison between the swirling strength, the $\Gamma$ functions, and the LAVD criteria on horizontal velocity fields extracted from observations.
An alternative approach has been put forward by \citet[][]{2021SoPh..296...17D} 
as they identify vortices directly from chromospheric filtergrams using the morphological winding-angle method.

A rigorous identification process is of fundamental importance to infer the statistical properties of vortical motions in the solar atmosphere and, consequently, their impact on chromospheric and coronal heating. The mentioned detection methods greatly helped to shape our current understanding, but they are prone to errors and misidentifications. 
As discussed above, weakly rotating vortices can be problematic for mathematical criteria such as the vorticity, the swirling strength, and the $\Gamma$ functions, especially in turbulent and dynamical flows such as the solar atmospheric ones. \citet[][]{2018ApJ...863L...2S} showed that an extra criterion, the $d$-criterion, should be used with LAVD/IVD methods and with the algorithm proposed by \citet[][]{2017A&A...601A.135K} in the presence of strong shear flows, which are typical of integranular regions. Moreover, the identification of the vortex center and boundaries, which is needed for a proper study of vortex dynamics and interaction, can be obtained with the $\Gamma$ functions and the LAVD/IVD methods only. In principle, the morphological method presented by \citet[][]{2021SoPh..296...17D} should be the preferred method because it takes the large-scale properties of the flow into account, but it is not well-suited for the analysis of numerical simulations.

Therefore, in this paper, we introduce a new vortex identification method based on a completely new technique. It takes into account the global features of the flow, as in the morphological methods, but also possesses the rigor of the mathematical criteria. 
Basically, it is a hybrid of the two classes presented above, and it possesses the ability of recognizing coherent vortical structures from the velocity field. Moreover, it allows for a precise characterization of the vortex center and boundaries and it is robust against shear flows and noise.

The paper is organized as follows. In Sect.\,\ref{sec:theoretical_backgroud} we present the mathematical criteria adopted in this work. 
The method and the automated algorithm are described in detail in Sect.\,\ref{sect:method}, while in Sect.\,\ref{sec:application_and_discussion} we test them on vortical and turbulent velocity fields and discuss the results.
Finally, we summarize and conclude in Sect.\,\ref{sec:conclusions}. 
The application of the method to numerical simulations and observations of the solar atmosphere is deferred to a follow-up paper.

%
%

\section{Theoretical Background}
\label{sec:theoretical_backgroud}

%
%

In this section we review some aspects of two of the most widely used mathematical criteria for vortical flows: the vorticity $\boldsymbol{\omega}$ and the swirling strength $\lambda$.
Moreover, we introduce a recently developed quantity called Rortex or Liutex, which is the one we adopt for the development of our method. 

%
%
\subsection{Vorticity}
\label{subsec:vorticity}
Vorticity is the classical quantity to describe local rotational motions in fluid mechanics and it is defined as the curl of the velocity field $\boldsymbol{v}$, 
\begin{equation}
    \boldsymbol{\omega} = \boldsymbol{\nabla} \times \boldsymbol{v}\,.\label{eq:vorticity}
\end{equation}
The direction of the vorticity vector indicates the orientation of the rotation according to the right-hand rule, while its norm is proportional to the intensity of the rotation. 

For a rotational vortex, that is a flow rotating in a rigid-body fashion around an axis, the norm of the vorticity vector is $\omega = |\boldsymbol{\omega}| = 2 \Omega$, where $\Omega$ is the angular velocity of the fluid. 
However, this simple relation between the vorticity norm and the attributes of vortical flows does not hold for more realistic vortex models in fluid dynamics, such as Lamb-Oseen or Burgers vortices. 
Therefore, we can not extract physical information about vortices directly from the vorticity. 
Moreover, the vorticity can assume large values also in the presence of shear flows and can therefore lead to false detections in non-rotating velocity fields.

%
%

\subsection{Swirling strength}
\label{subsec:swirlingstrength}
A suitable solution to the shear flow problem is given by the swirling strength criterion, $\lambda$, proposed by \citet{1999JFM...387..353Z}. 
It is computed through the eigen analysis of the velocity gradient tensor, $\mathcal{U} = \nabla \boldsymbol{v}$, that is the Jacobian matrix of the velocity field.
If the flow is locally rotating, the velocity gradient tensor can be diagonalized as,
\begin{align}
  \mathcal{U} =  
    \underbrace{\vphantom{\begin{bmatrix}
		                  \lambda_{\rm r} & 0 &  0\\
		                  0 & \lambda_{\rm +}& 0 \\
		                  0 & 0 & \lambda_{\rm -}  
		                  \end{bmatrix}} 
    \left[ \boldsymbol{u}_{\rm r}, \boldsymbol{u}_{\rm +}, \boldsymbol{u}_{\rm -}\right]
    }_{\textstyle\mathcal{P}}
	\underbrace{%
	\begin{bmatrix}
      \lambda_{\rm r} & 0 &  0\\
      0 & \lambda_{\rm +}& 0 \\
      0 & 0 & \lambda_{\rm -}  
    \end{bmatrix}}_{\textstyle\Lambda}
    \underbrace{\vphantom{\begin{bmatrix}
		                  \lambda_{\rm r} & 0 &  0\\
		                  0 & \lambda_{\rm +}& 0 \\
		                  0 & 0 & \lambda_{\rm -}  
		                  \end{bmatrix}} 
	\left[ \boldsymbol{u}_{\rm r}, \boldsymbol{u}_{\rm +}, \boldsymbol{u}_{\rm -}\right]^{-1}
	}_{\textstyle\mathcal{P}^{-1}}\,,  \label{eq:U_decomposition}
\end{align}
where $\lambda_{\pm} = \lambda_{\rm cr} \pm  \mathrm{i}\lambda_{\rm ci} \in \mathbb{C}$
and $\lambda_{\rm r} \in \mathbb{R}$ are the eigenvalues of $\mathcal{U}$ that form the eigenvalue matrix $\Lambda$, while $\boldsymbol{u_{\pm}}$ and $\boldsymbol{u_{\rm r}}$ are the corresponding eigenvectors that form the change of basis matrix $\mathcal{P}$.
We define the swirling strength $\lambda$ as twice the imaginary part of the complex eigenvalues, 
\begin{equation}
    \lambda = 2\lambda_{\rm ci}\,. \label{eq:swirling_strength}
\end{equation} 
Moreover, the normalized eigenvector\footnote{Such that its norm is $1$.} associated with the real eigenvalue, $\boldsymbol{u}_{\rm r}$, defines the rotation axis\footnote{The eigenvector $\boldsymbol{u}_{\rm r}$ is actually defined up to a $\pm$ sign. 
To have the eigenvector pointing in the direction of the rotation according to the right-hand rule, 
one must check the orientation of the orthogonal basis defined by the three eigenvectors. The reader can refer to \citet[][]{2020A&A...639A.118C} for more details. }. 
We can then define the swirling strength vector as $\boldsymbol{\lambda} = \lambda \boldsymbol{u}_{\rm r}$, which carries the information about the strength and the direction of rotation. For a simple rotational vortex, the swirling strength vector is identical to the vorticity vector. 
For more information about the swirling strength vector the reader can refer to \citet{2020A&A...639A.118C}. 

The swirling strength criterion can be enhanced by considering the inverse spiraling compactness $\zeta = \lambda_{\rm cr}/\lambda_{\rm ci}$, where $\lambda_{\rm cr}$ is the real part of the complex eigenvalues of $\mathcal{U}$ \citep[][]{2005JFM...535..189C}. 
The inverse spiraling compactness is a measure of the radial distance traveled by a test particle in the flow during one orbital period. 
It can be shown that the radial distance $r_{\rm n}$ of a particle spiraling in the vortex plane evolves as, 
\begin{equation}
    r_{\rm n} = r_0 \exp{\left(2\pi n \zeta \right)}\,, \label{eq:spiral_compactness}
\end{equation}
where $r_0$ is the initial radius and $n$ is the number of orbits \citep[][]{2005JFM...535..189C}. 
If $\zeta = 0$, then the flow is perfectly circular, while if $\zeta$ is positive (negative) the flow is spiraling outward (inward). 
Large (positive or negative) values of $\zeta$ indicate very low spiraling compactness. 
Therefore, vortex regions identified by the swirling strength should be discarded if $|\zeta|$ is too large, since in these cases the orbital component of the velocity field is negligible compared to the radial one and the rotational motion is barely perceivable. 
In practice, we set $\lambda = 0$ wherever $ \kappa_{\zeta} > \zeta > \delta_{\zeta}$, where $\kappa_{\zeta}$ and $\delta_{\zeta}$ are free parameters that can be adjusted according to the compactness required. Typical values of these parameters are $-\kappa_{\zeta} \sim \delta_{\zeta} \lesssim 1$.

The swirling strength solves the problem related to pure shear flows, but it still fails to predict the angular velocity of more complex and realistic vortices. 
Indeed, the swirling strength criterion can not discern between the rigid-body rotational component of a flow and intrinsic shears, which are present in differentially rotating flows for example (see Sect.\,\ref{subsec:Rortex}). 
This means that basic information about vortices, such as their period of rotation and their size, cannot be directly inferred from the simulation data using the swirling strength criterion,
as we show in Sect.\,\ref{subsec:comparing_criteria}.

%
%

\subsection{Rortex} 
\label{subsec:Rortex}
The Rortex criterion, also called Liutex, is a state of the art mathematical quantity, proposed by \citet{2018JFM...849..312T} and \citet[][]{2018PhFl...30c5103L}, 
which allows for a precise quantification of the local rigid-body rotational part of the flow alone. Indeed, in complex but realistic vortex models, an intrinsic shear component is usually present besides the rigid-body rotational one. An example of such a model is the Lamb-Oseen vortex (see Sect.\,\ref{subsec:comparing_criteria}). The intrinsic shear component blends with the purely rotational one in the elements and eigenvalues of the velocity gradient tensor. Since the vorticity and the swirling strength cannot discern between these two components of the flow, both can be contaminated by the presence of an intrinsic shear. 

To precisely represent the fluid curvature at a point, \citet[][]{2016SCPMA..59h..22L} 
proposed to decompose the vorticity vector $\boldsymbol{\omega}$ into a purely rotational part $\boldsymbol{R}$ and a non-rotational or shear component $\boldsymbol{S} = \boldsymbol{\omega} - \boldsymbol{R}$. The direction of $\boldsymbol{R}$ represents the rotation axis, while its norm is proportional to the angular velocity of the flow. For a purely rigid-body rotational flow, $\boldsymbol{R} = \boldsymbol{\lambda} = \boldsymbol{\omega}$ at any point because there are no shears in such a flow. However, a series of tedious matrix transformations are in principle necessary to compute the vectors $\boldsymbol{R}$ and $\boldsymbol{S}$ and therefore extract the rotational component of the flow.

As shown by \citet[][]{2019PhFl...31i5102X}, 
for any velocity gradient tensor $\mathcal{U} = \nabla \boldsymbol{v}$ with one real and two complex conjugated eigenvalues, there exists a transposed quasi-triangular matrix $\mathcal{V}$ such that,
\begin{equation}
    \mathcal{U} = \mathcal{Q}\,\mathcal{V}\,\mathcal{Q}^{\rm T}\,,\label{eq:schur_dec_vgt}
\end{equation}
where $\mathcal{Q}$ is an orthogonal matrix representing a rotation operator in three-dimensional space and $\mathcal{V}$ can be written as,
\begin{equation}
    \mathcal{V} = \begin{bmatrix}
        \lambda_{\rm cr}   & -\phi            & 0 \\
        \phi + \varepsilon & \lambda_{\rm cr} & 0 \\
        \xi                & \nu              & \lambda_{\rm r} 
    \end{bmatrix}\,. \label{eq:schur_form_vgt}
\end{equation}
The matrix $\mathcal{V}$ represents the velocity gradient tensor $\mathcal{U}$ in a rotated reference frame with the rotation axis parallel to the new $\hat{z}$-axis. We recall that the rotation axis is given by the normalized eigenvector associated with the real eigenvalue $\lambda_{\rm r}$ of $\mathcal{U}$, as described in Sect.\,\ref{subsec:swirlingstrength}. 

The various components of $\mathcal{V}$ represent rigid-body rotation $(\phi)$, stretching and compressing components $(\lambda_{\rm cr},\lambda_{\rm r})$, and shearing parts $(\varepsilon, \xi, \nu)$. In particular, the $\varepsilon$ coefficient is mixed with the purely rotational component $\phi$ and it is responsible for the intrinsic shear. 
From Eq.\,(\ref{eq:schur_dec_vgt}) it becomes clear why the swirling strength can be biased by the presence of intrinsic shear flows, as the eigenvalues of the matrix $\mathcal{V}$ are contaminated by the $\varepsilon$ coefficient. Hence, the swirling strength directly depends on the strength of intrinsic shears and it can not be used to measure the strength of the rigid-body rotational part of the flow.

To this aim, the Rortex criterion is defined as,
\begin{equation}
    R = 2\phi\,. \label{eq:Rortex_def}
\end{equation}
By definition, it computes the strength of the rotational part of the flow alone. When there are no intrinsic shears (i.e., $\epsilon = 0$), one can show that the Rortex criterion is equivalent to the swirling strength by diagonalizing Eq.\,(\ref{eq:schur_form_vgt}). Furthermore, one can compute the rotational part $\boldsymbol{R}$ of the vorticity vector $\boldsymbol{\omega}$ by multiplying the Rortex criterion with the rotation axis given by the normalized eigenvector $\boldsymbol{u}_{\rm r}$,
\begin{equation}
    \boldsymbol{R} = R\,\boldsymbol{u}_{\rm r}\,. \label{eq:Rortex_vector}    
\end{equation} 
This vector is referred to as Vortex or Rortex vector,
and it allows for a three-dimensional characterization of the vortical flow. 

To compute the Rortex criterion one would have to find the orthogonal matrix $\mathcal{Q}$ such that the rotated local velocity gradient tensor $\mathcal{V}$ takes the form of Eq.\,(\ref{eq:schur_form_vgt}) \citep[see ][for a guided procedure]{2018PhFl...30h5107G}. 
Fortunately, \citet[][]{2019JHyDy..31..464W} and \citet[][]{2019PhFl...31i5102X} derived a simple formula to compute the Rortex criterion without having to dive into complex and computationally expensive matrix calculations. Accordingly,
\begin{equation}
    R = 2\phi = \boldsymbol{\omega}\cdot\boldsymbol{u}_{\rm r} - \sqrt{(\boldsymbol{\omega}\cdot\boldsymbol{u}_{\rm r})^2 - \lambda^2}\,,\label{eq:Rortex}
\end{equation}
where $\boldsymbol{\omega}$ is the vorticity vector (Sect.\,\ref{subsec:vorticity}) and $\lambda = 2 \lambda_{\rm ci}$ is the swirling strength criterion (Sect.\,\ref{subsec:swirlingstrength}). Here, the sign of the normalized eigenvector $\boldsymbol{u}_{\rm r}$ is chosen such that $\boldsymbol{\omega}\cdot\boldsymbol{u}_{\rm r} > 0$. 
Also, the Rortex criterion can be enhanced in the same way \citet{2005JFM...535..189C} 
did it for the swirling strength. Hence we set $R=0$ wherever $\kappa_{\zeta} > \zeta > \delta_{\zeta}$, with $\kappa_{\zeta}$ and $\delta_{\zeta}$ chosen thresholds (see Sect.\,\ref{subsec:swirlingstrength}). 

The Rortex criterion allows for a precise quantification of the rotational strength of a vortex flow. Indeed, it is the only quantity not being contaminated by the intrinsic shear component $\varepsilon$ in the velocity gradient tensor. Therefore, it is the only reliable quantity for the extraction of physical information about a vortex, such as the rotational period and the curvature radius.  

%
%

\subsection{Comparing criteria} 
\label{subsec:comparing_criteria}
\begin{figure*}
	\centering
	\resizebox{\hsize}{!}{\includegraphics{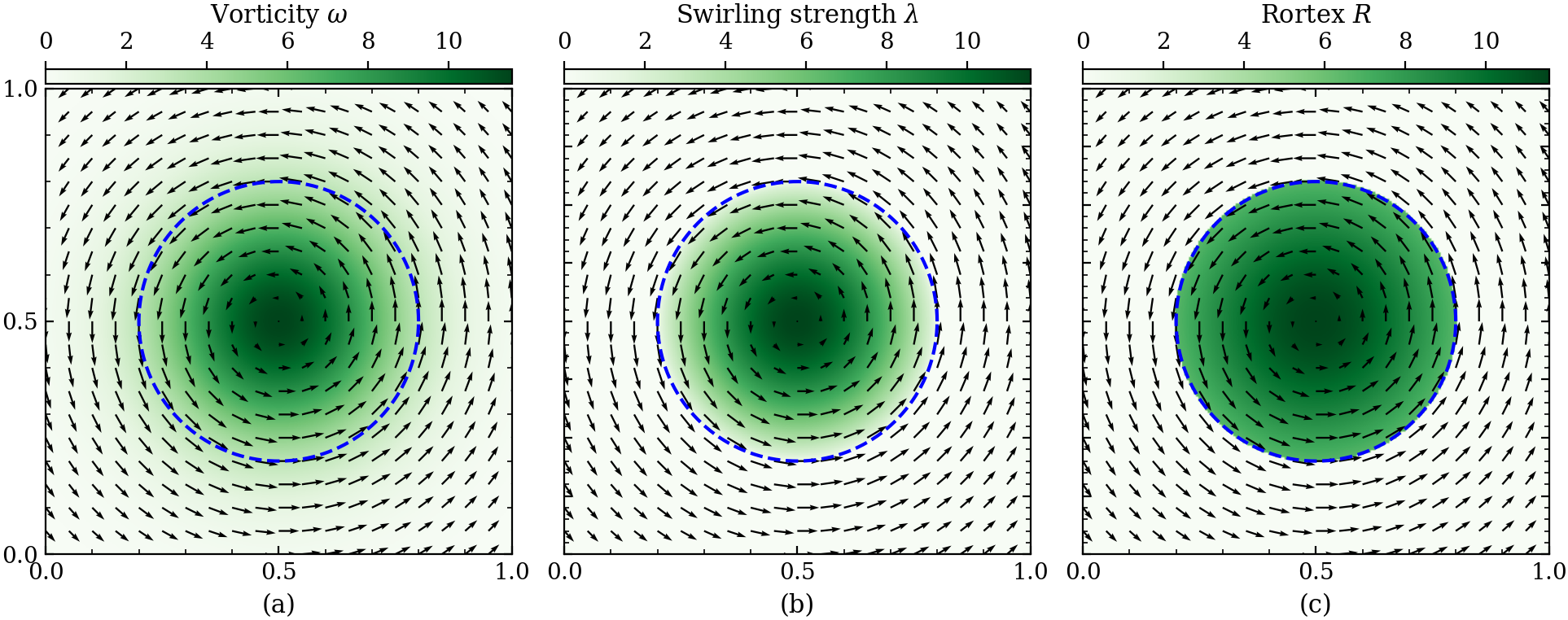}}
	\caption{Comparison between (\textit{a}) the vorticity, (\textit{b}) the swirling strength, and (\textit{c}) the Rortex criteria for a Lamb-Oseen vortex. Arrows indicate the velocity vector field of the Lamb-Oseen vortex and their length is proportional to the magnitude of the flow. The blue-dashed circles denote the boundary of the vortex defined by $r = r_{\rm max}$.}
	\label{fig:compare_crit}
\end{figure*}
\begin{figure}
	\centering
	\resizebox{\hsize}{!}{\includegraphics{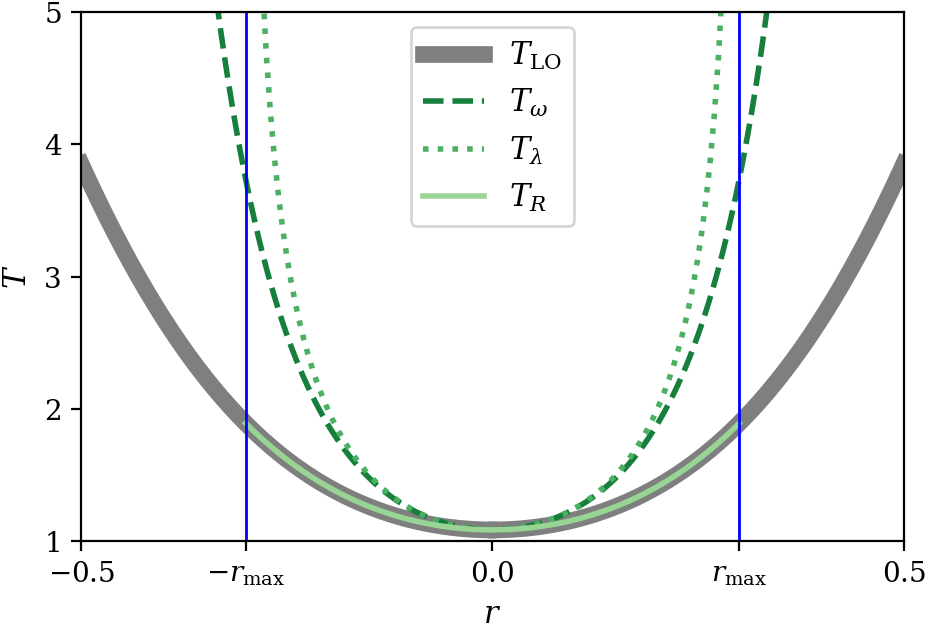}}
	\caption{Comparison of the estimated rotational period $T$ derived from the vorticity, the swirling strength, and the Rortex criteria  with the true rotational period for a Lamb-Oseen vortex. Blue vertical lines denote the boundary of the vortex defined by $r=r_{\rm max}$. The thick gray curve indicates the true rotational period $T_{\rm LO}$ computed from the analytical formula Eq.\,(\ref{eq:lamb_oseen}). }
	\label{fig:compare_crit_period}
\end{figure}

To prove the reliability of the Rortex criterion, we compare the three mathematical criteria presented in Sects.\,\ref{subsec:vorticity}, \ref{subsec:swirlingstrength}, and \ref{subsec:Rortex} by applying them to a Lamb-Oseen vortex model. 
Lamb-Oseen vortices are analytical but realistic vortex models and are defined through the following velocity field in polar coordinates,
\begin{equation}
    v_{\theta} = v_{\rm max}\left(1 + \frac{1}{2\alpha} \right)\frac{r_{\rm max}}{r}\left[1 - \exp{\left(-\alpha\frac{r^2}{r_{\rm max}^2}\right)}\right]\,,~~ v_r = 0 \label{eq:lamb_oseen}
\end{equation}
where $\alpha = 1.256$ establishes that $r_{\rm max}$ is the radius at which the rotational velocity $v_{\theta}$ is maximal and reaches the value $v_{\rm max}$. 
We then define $r_{\rm max}$ to be the boundary of the vortex. In the following,  we set $v_{\rm max}=1$ and $r_{\rm max}=0.3$.

The velocity field corresponding to the vortex model presented above and centered in $(x,y) = (0.5,\,0.5)$ is shown in the three panels of Fig.\,\ref{fig:compare_crit} with arrows proportional to the local strength of the velocity field, while the blue dashed lines represent the boundary of the vortex given by $r=r_{\rm max}=0.3$.
The panels show, from left to right, the maps of the vorticity $\omega$, the swirling strength $\lambda$, and the Rortex $R$. 
As we can see, the three criteria identify the presence of local curvature in the flow. 
The swirling strength and the Rortex are non-zero only inside the boundary of the vortex, while vorticity shows positive values also for $r>r_{\rm max}$. 
Moreover, the three criteria show different values in the vortical region. 

To know which criteria is the most appropriate to measure the characteristics of such a vortical flow, we compare the rotational period $T$ estimated with the vorticity, the swirling strength, and the Rortex at each point in the grid. Given the symmetry of the problem, the rotational period $T$ depends only on the radius $r$. We assume that the angular velocity $\Omega$ and these quantities are related by $c = 2\Omega_{c}$, where $c = \omega, \lambda, R$. Then, we can estimate the rotational period as,
\begin{equation}
    T_{c} = \frac{2\pi}{|\Omega_{c}|} = \frac{4\pi}{|c|}\,,  \label{eq:period_criteria}
\end{equation}
and compare it to the true period derived from the analytical formula Eq.\,(\ref{eq:lamb_oseen}), that is,
\begin{equation}
    T_{\rm LO} = \frac{2\pi}{|\Omega_{\rm LO}|} = \frac{2\pi r}{|v_{\theta}|}\,, \label{eq:period_analytical}
\end{equation}
since $\Omega_{\rm LO} = v_{\theta}/r$. 
Note that $|c|$ in Eq.\,({\ref{eq:period_criteria}}) can be a function of radius as is $\Omega_{\rm LO}$.

Figure \ref{fig:compare_crit_period} shows the radial profiles of such estimated rotational periods against the true value $T_{\rm LO}$. 
The only criterion able to correctly estimate the true rotational period of the vortex flow as a function of the radius is the Rortex. This is because intrinsic shear contaminate the purely rotational component of Lamb-Oseen vortices. 
Therefore, the Rortex criterion is the optimal quantity to extract precise information about the rotational characteristics of this kind of vortices. 
This conclusion can be drawn also analytically and remains valid for many other vortex models,
as we show in Appendix \ref{app:analytical_comparison}.

In principle, one can define a vortical region as a collection of connected grid cells where $R \neq 0$. However, this definition only takes into account the local properties of the flow and lacks of a more global perspective. Since a vortex is a coherent structure that has a spatial extension, it is important to consider the relation between all the fluid parcels that compose it.
That is why, in the next section, we combine the Rortex criterion with a morphological method. 

%
%

\section{Method}
\label{sect:method}

In this section, we present a new vortex identification method. 
As stated above, this new method is based on both morphological and mathematical criteria. 
More precisely, it is an extension of the curvature center method proposed by \citet{Sadar99}, where we adopt the Rortex criterion to compute the coordinates of the curvature center. 
In this way, we solve the main flaw of the curvature center method, that is it being based on the curvature of streamlines of the flow. 
Our method takes the intuitive idea of the curvature center method and combines it with a state of the art mathematical criterion (the Rortex criterion) for a precise estimation of the curvature center position for all points where the velocity field is curved.

%
%

\subsection{Estimated vortex center map}  
\label{subsec:EVCmap}

The basic idea of the method presented in this paper is to compute a map with all the ``Estimated Vortex Centers'' (EVC), that is, the curvature center points, from any parcel of flow presenting a rigid-body rotational component.
From here on, we dub this map the EVC map. Whenever a vortex with a rigid-body rotational component\footnote{Every naturally occurring vortex is expected to have a rigid-body rotational component in its flow. The only known vortex model that does not have it is the peculiar case of the irrotational vortex.} 
is present in the velocity field, a high density (or a cluster) of EVC points will show up in this map, since all points belonging to the vortex share a common curvature center. 
If instead the flow is characterized by incoherent random local curvatures only, the EVC points will be scattered in the domain and they will not form clusters.
The reliability of such a map, and therefore of the method, depends on how precisely the vortex center can be estimated. 
In the following, we show how to mathematically estimate the center of curvature, that is, how to calculate the EVC map.

For simplicity, we consider only a two-dimensional flow, but this methodology can be extended to three dimensions. 
Given a generic velocity field $\boldsymbol{v} = (v_x, v_y)$ defined in a Cartesian grid with coordinates $\boldsymbol{x} = (x,y)$, the vorticity $\boldsymbol{\omega}$, the velocity gradient tensor $\mathcal{U}$, the real eigenvector $\boldsymbol{u}_{\rm r}$, the swirling strength $\lambda$, and the Rortex $R$ can be computed for all the points in the grid according to Eqs.(\ref{eq:vorticity}), (\ref{eq:swirling_strength}), and (\ref{eq:Rortex}). 
In the following we focus on a generic point with coordinates $\boldsymbol{x}_0 = (x_0, y_0)$, Rortex value $R_0 \neq 0$, and velocity $\boldsymbol{v}_0 = (v_{x,0}, v_{y,0})$. 
Moreover, we assume the velocity gradient tensor in $\boldsymbol{x}_0$, $\mathcal{U}|_{\boldsymbol{x}_0}$, to be known. 
Then, the EVC of the point $\boldsymbol{x}_0$ can be computed based on two quantities: 
the curvature radius $r$, that is the distance between the point $\boldsymbol{x}_0$ and the center of curvature, and the radial direction $\boldsymbol{e}_{\rm r}$, that is the unit vector pointing from the point $\boldsymbol{x}_0$ to the center of curvature.

%
\begin{figure}
	\centering
	\includegraphics{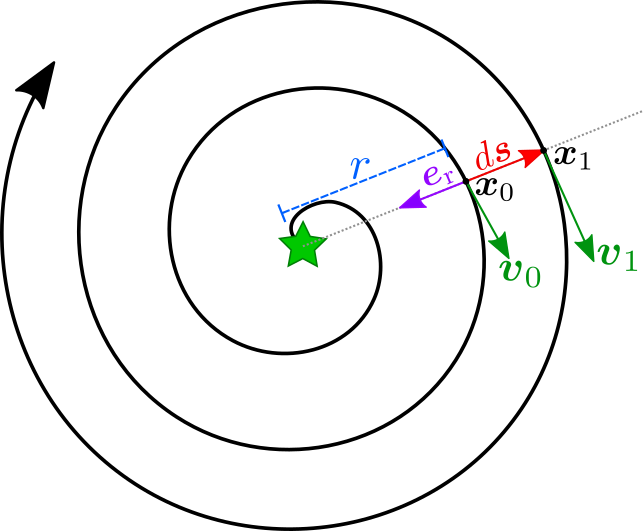}
	\caption{Representation of a vortical velocity streamline. The velocity field at two points $\boldsymbol{x}_0$ and $\boldsymbol{x}_1$ is shown by green arrows and denoted $\boldsymbol{v}_0$ and $\boldsymbol{v}_1$, respectively. The two points are relatively close to each other and the vector joining them is labeled $d\boldsymbol{s}$. The point $\boldsymbol{x}_0$ is at distance $r$ from the vortex center that is marked with a star. The purple unit vector $\boldsymbol{e}_{\rm r}$ represents the radial direction. }
	\label{fig:sketch}
\end{figure}
%

%
%

\subsubsection{The radial direction}
To estimate the radial direction in a generic point $\boldsymbol{x}$, the naive idea would be to take the vector perpendicular to the velocity field in $\boldsymbol{x}$.
However, if the flow in this point has a radial velocity component or shear, the estimated vector will deviate from the true radial direction. 
Moreover, the perpendicular vector defines a plane in three dimensions, and one would need to rely on other quantities to determine the direction. 
Therefore, we adopt a method based on the Taylor expansion of the velocity field at any point $\boldsymbol{x}$.

Let us consider two points, $\boldsymbol{x}_0$ and $\boldsymbol{x}_1$, which we assume to be along the radial direction of the vortex and relatively close to each other.
The scenario is presented in Fig.\,\ref{fig:sketch}. We can expand the velocity field
in $\boldsymbol{x}_1$, $\boldsymbol{v}_1$, around $\boldsymbol{x}_0$ as,
\begin{equation}
    \boldsymbol{v}_1 = \boldsymbol{v}_0 + \boldsymbol{\nabla}\boldsymbol{v} |_{\boldsymbol{x}_0}\, d\boldsymbol{s} + \mathcal{O}(|d\boldsymbol{s}|^2)\,, \label{eq:taylor_expansion}
\end{equation}
where $d\boldsymbol{s} = \boldsymbol{x}_1 - \boldsymbol{x}_0$ and $\boldsymbol{\nabla}\boldsymbol{v}|_{\boldsymbol{x}_0} = \mathcal{U}|_{\boldsymbol{x}_0}$ is the velocity gradient tensor computed in $\boldsymbol{x}_0$. If $|d\boldsymbol{s}| = |\boldsymbol{x}_1 - \boldsymbol{x}_0|$ is sufficiently small, we can neglect the higher order terms. 
Moreover, the two points of interest are aligned along the radial direction by construction, hence $d\boldsymbol{s}$ is proportional to $\boldsymbol{e}_{\rm r}$.

Both the velocity gradient tensor $\mathcal{U}|_{\boldsymbol{x}_0}$ and the distance vector $d\boldsymbol{s}$ are Galilean invariant, therefore we can safely set $\boldsymbol{v}_1 = 0$. Moreover, if $\boldsymbol{x}_0$ belongs to a vortical region, then $\mathcal{U}|_{\boldsymbol{x}_0}$ is diagonalizable \citep[see, e.g.,][]{2020A&A...639A.118C} 
and therefore also invertible. Hence, we can write Eq.\,(\ref{eq:taylor_expansion}) as,
\begin{equation}
    d\boldsymbol{s} = \mathcal{U}^{-1}|_{\boldsymbol{x}_0} \boldsymbol{v}_0\,, \label{eq:estimated_direction}
\end{equation}
and, since $d\boldsymbol{s} \propto \boldsymbol{e}_{\rm r}$, we have a simple formula to compute the radial direction. It is important to notice that Eq.\,(\ref{eq:estimated_direction}) alone only yields information about the straight line connecting the center of curvature and the point $\boldsymbol{x}_0$. Therefore we also need to compute the radius of curvature to find the coordinates of the EVC.

%
%

\subsubsection{The curvature radius}
\label{subsubsec:curvature_radius}

\begin{figure*}
	\centering
	\resizebox{\hsize}{!}{\includegraphics{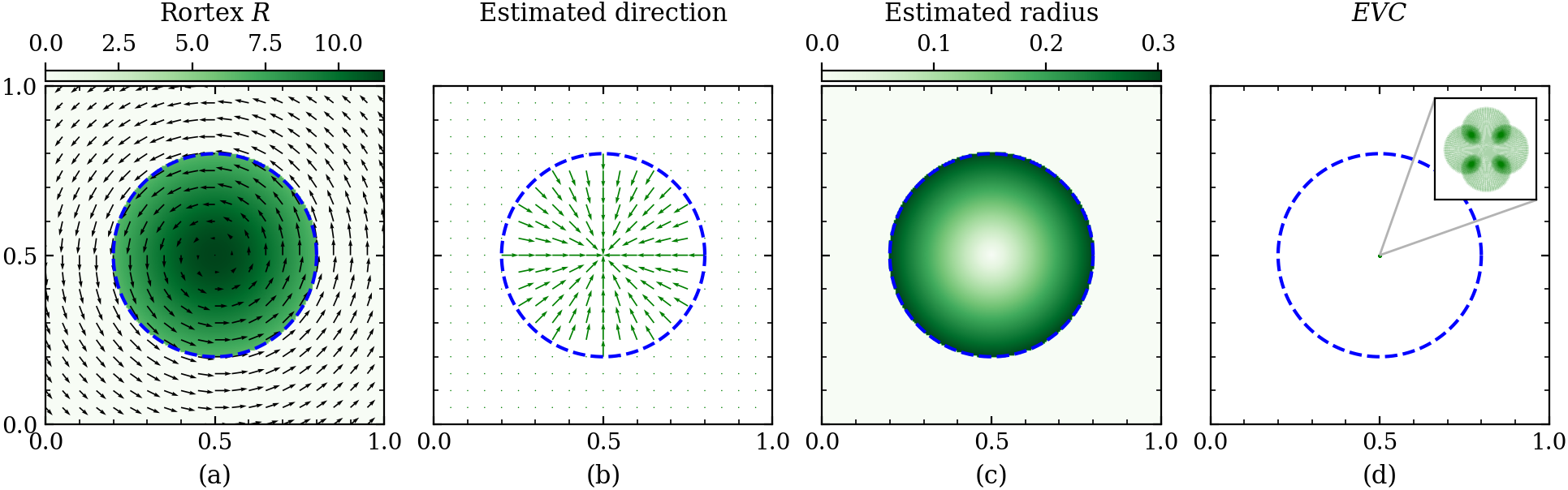}}
	\caption{Demonstration of the application of the EVC method to the Lamb-Oseen vortex of Fig.\,\ref{fig:compare_crit}. Steps are as follows: (\textit{a}) computation of the Rortex criterion, (\textit{b}) estimation of the radial direction, (\textit{c}) estimation of the curvature radius, and (\textit{d}) computation of the EVCs. The blue-dashed circles denote the boundary of the vortex defined by $r = r_{\rm max}$.}
	\label{fig:demo_lamboseen}
\end{figure*}

In Sect.\,\ref{subsec:comparing_criteria} we have accurately estimated the rotation period of the Lamb-Oseen
vortex using the Rortex criterion $R$, that is $T_{\rm R} = \frac{4\pi}{|R|}$. Therefore, the curvature radius at the point $\boldsymbol{x}_0$ can be estimated by combining Eq.\,(\ref{eq:period_criteria}) with the rotational velocity as,
\begin{equation}
    \frac{4\pi}{|R_0|} = \frac{2\pi r}{|v_{\theta}|}\,, \label{eq:combining_periods}
\end{equation}
where $v_{\theta}$ is the tangential component of the velocity field. Assuming $v_{\theta} = |\boldsymbol{v}_0|$, where $|\boldsymbol{v}_0|$ is the norm of the velocity field in $\boldsymbol{x}_0$, Eq.\,(\ref{eq:combining_periods}) can be expressed as
\begin{equation}
    r = \frac{2 |\boldsymbol{v}_0|}{|R_0|}\,, \label{eq:estimated_radius}
\end{equation}
and the curvature radius can be estimated based alone on the norm of the velocity field and the Rortex criterion in $\boldsymbol{x}_0$. 

The assumption that $v_{\theta} = |\boldsymbol{v}_0|$ is valid if the flow has no radial component. In more realistic cases, Eq.\,(\ref{eq:estimated_radius}) still holds as a first approximation since the tangential component of the velocity field must dominate over the radial one for the flow to be part of a vortex system. The estimation can probably be improved by considering the stretching, compression, and shearing components of the velocity gradient tensor, as seen in Sect.\,\ref{subsec:Rortex}.

Finally, combining Eqs.\,(\ref{eq:estimated_direction}) and (\ref{eq:estimated_radius}), we are left with two points that are at distance $r$ from $\boldsymbol{x}_0$ and on the straight line defined by $\boldsymbol{e}_r$.
To decide which one to select as EVC, we check the orientation of the local curvature given by the sign of the swirling strength\footnote{In three dimensions, one would use the direction of the swirling strength vector given by the real eigenvector $\boldsymbol{u}_{\rm r}$.}. 
If the swirling strength is positive, that is the flow is curved in the anti-clockwise direction, then the correct point is the one on the left with respect to the direction of the flow. 
If the sign of the criterion is negative, then the flow is is curved in the clockwise direction and the point on the right is to be selected. 

In Fig.\,\ref{fig:demo_lamboseen} we apply the method, step by step, to the same Lamb-Oseen vortex of Sect.\,\ref{subsec:comparing_criteria}. First, we compute the Rortex criterion on the whole grid (panel a). Then, wherever $R \neq 0$, we estimate the radial direction vectors (panel b) and the curvature radii (panel c) according to Eq.\,(\ref{eq:estimated_direction}) and Eq.\,(\ref{eq:estimated_radius}), respectively.  
Finally, we combine the results as discussed above and obtain the EVC map (panel d). 
As we can see, the method estimates very precisely the position of the vortex center as all the EVCs are clustered around it. 
The mean error of the estimation is of the order of the grid cell size, and it is mainly due to discretization errors in the numerical computation of the velocity field derivatives. 

In more complicated scenarios, where turbulence or noise can distort the velocity field at the grid cell level, the accuracy of this method will not be as impressive as in Fig.\,\ref{fig:demo_lamboseen}. In such cases, a ``multiple stencil'' approach can be used to improve the method's robustness. The idea is to compute and assemble together multiple EVC maps to help discern between true clusters and noisy points. To obtain different EVC maps, one computes the necessary derivatives using different stencils of grid cells
but maintaining the same order of accuracy. 

An ordinary, second order, central finite difference derivative uses three adjacent grid cells, which locations on the grid can be arbitrarily defined as -1,0, and 1. We dub this derivative a stencil-1 derivative, because the distance between the used grid cells is 1 grid cell size. A stencil-2 derivative of the same type operates over grid cells separated by two grid cell sizes, that is at locations -2,0, and 2. Both derivatives are second order in accuracy, but the stencil-1 derivative is better suited to capture variations in the velocity field at the grid cell size level, while larger stencils can be used to infer large-scale variations. By combining EVC maps computed with different stencils, one can discern small-scale perturbations from actual vortical structures and improve robustness against noise.

%
%

\subsection{The algorithm} 
\label{subsec:algorithm}

The method presented in Sect.\,\ref{subsec:EVCmap} can be used as it is for a manual identification of vortices by searching for clusters of EVC points. 
However, when it comes to analyze large and turbulent velocity fields, as for example the ones resulting from numerical simulations of convection, it can be useful to have an algorithm that processes the EVC map and automatically finds the vortices in it. 
In the following, we present the automated detection algorithm based on the EVC map method.

\subsubsection{Clustering method}
\label{subsubsec:clustering_method}

Given the nature of the EVC map, we opt for a clustering algorithm. 
The idea is to classify EVCs into categories, or clusters, based on their distance relatively to each other. 
Each cluster represents a vortex, for which one can compute different properties. A similar approach has been adopted by \citet[][]{2021SoPh..296...17D} 
to automatically identify vortices from clusters of curved streamlines. 

For this work, we chose to use the clustering by fast search and find of density peaks (CFSFDP) proposed by \citet{2014Sci...344.1492R}. 
The basic assumption of the algorithm is that cluster centers can be defined as locations characterized by relatively high local densities of datapoints (here EVCs) that are at relatively large distance from other datapoints with higher local density.
Mathematically, these two criteria rely on the computation of two quantities: the local density of datapoints $\rho$ and the spacing from other datapoints of higher local density $\delta$. 
Both of them depend only on the Euclidean distance $d_{ij}$ between datapoints $i$ and $j$, that is, the distance between two EVCs $i$ and $j$. Therefore, no human interaction is required and the algorithm has shown high performance with different datasets and robustness against noise \citep[][]{2014Sci...344.1492R}. 

The local density $\rho_i$ relative to each datapoint $i$ is computed as,
\begin{equation}
    \rho_i = \sum_j \chi( d_{ij},\, d_c ) \, , \label{eq:def_density}
\end{equation}
where $\chi$ is a kernel that depends on the Euclidean distance $d_{ij}$ to neighboring points $j$ and on an arbitrary cutoff distance $d_c$. The diagonal terms $d_{ii}$ of the distance matrix, that is, the self-distances, are $0$ by definition.
The role of the kernel $\chi$ is to evaluate the proximity of two datapoints. 
\citet{2014Sci...344.1492R} 
use a cutoff kernel,
\begin{equation}
    \chi (d_{ij},\, d_c) = \left\{ \begin{aligned} 
  1,& ~~ \text{if}\, d_{ij} - d_c < 0\,, \\
  0,& ~~ \text{else}\,,
\end{aligned} \right. \label{eq:density_kernel_1}
\end{equation}
so that the density $\rho$ is the number of datapoints (EVCs) within a hyper-sphere of radius $d_c$.
Another possibility is to use a Gaussian kernel of the type,  
\begin{equation}
    \chi (d_{ij},\, d_c) = \exp{\left( - \frac{d_{ij}^2}{d_c^2}\right)}\,. \label{eq:density_kernel_2}
\end{equation}
The performance of the different kernels depend on the choice of the parameter $d_c$. As a rule of thumb, \citet{2014Sci...344.1492R} 
suggest to choose the cutoff parameter $d_c$ such that the average number of neighbor datapoints, that is those for which $d_{ij} < d_c$, is around 1 to 2\% of the total number of datapoints. \citet{2016Neucom...208.210M} 
propose instead a nonparametric method to estimate the cutoff distance. 

The spacing $\delta_i$ criterion is given by the distance between the generic datapoint $i$ and the closest datapoint $j$ with a higher local density, that is $\rho_j > \rho_i$. Mathematically, this statement can be written as, 
\begin{equation}
    \delta_i = \min_{j\,:\,\rho_j > \rho_i}(d_{ij})\,.  \label{eq:distance}
\end{equation}
This formulation ensures that only datapoints characterized by a local maximum in local density have a $\delta$ criterion considerably larger than the typical distance between neighboring datapoints. 
Therefore, datapoints with large density $\rho$ and spacing $\delta$ are potential cluster centers. Equation\,(\ref{eq:distance}) is applicable to all datapoints but the one with maximum local density. For this datapoint, the $\delta$ criterion is usually set to be $\delta = \max{(d_{ij})}$.

\subsubsection{Grid and vortex adaptation}
\label{subsubsec:grid_adaptation}

\begin{figure}
	\centering
	\resizebox{\hsize}{!}{\includegraphics{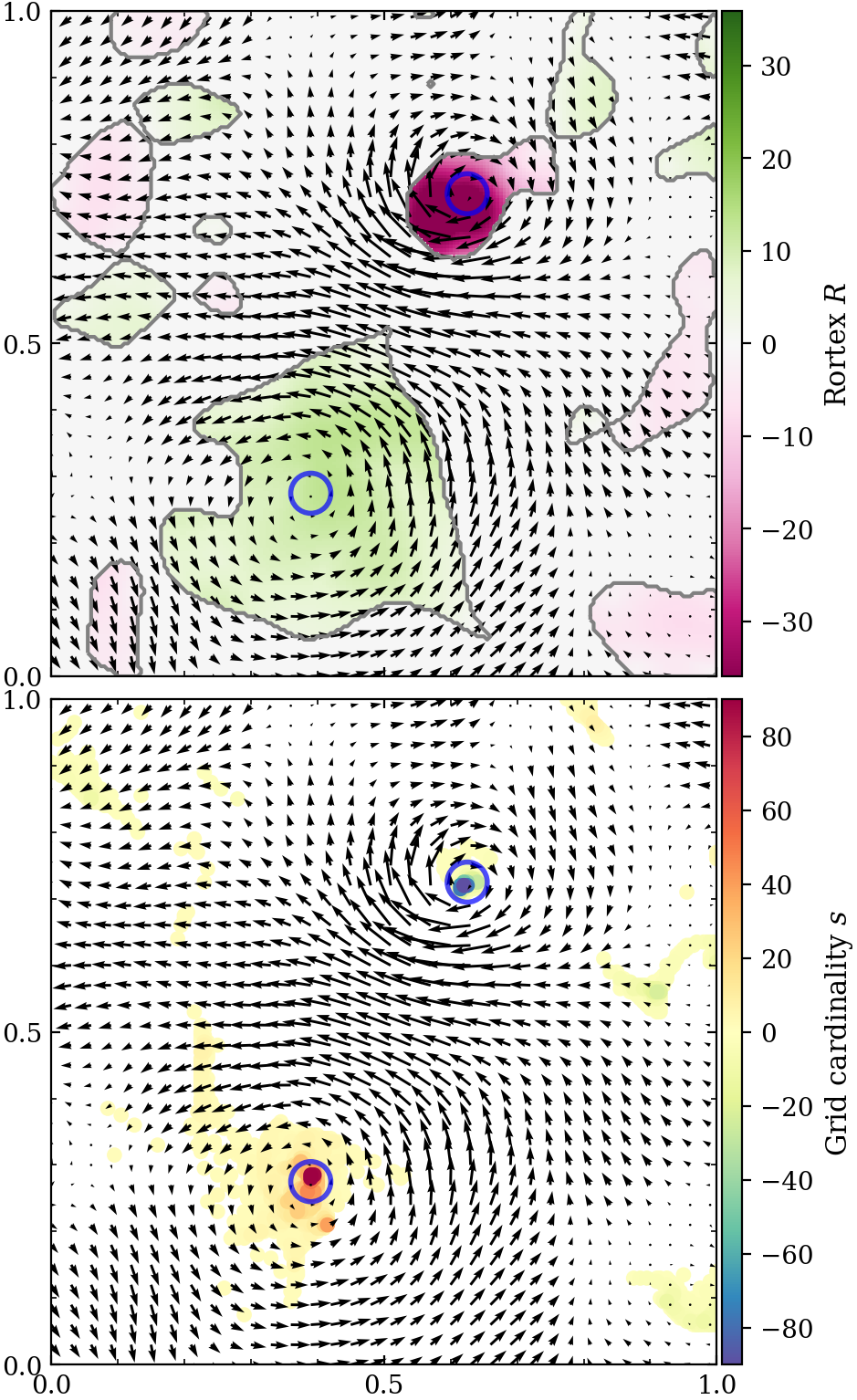}}
	\caption{Application of the grid-adapted EVC method to an artificial flow composed of two Lamb-Oseen vortices of opposite orientation superposed by a Gaussian noise signal. \textit{Top panel}: Rortex map. \textit{Bottom panel}: G-EVC map. The cardinality $s$ accounts for the number of EVC points which coordinates fall in the same grid cell. The blue circles show the approximate location of the two vortex centers. 
    The regions where $R \neq 0$ are outlined by a gray contour in the Rortex map.
	}
	\label{fig:demo_doublevortex}
\end{figure}

The CFSFDP method is mainly limited by its computational cost, which scales as $\mathcal{O}(N_{\rm P}^2)$, where $N_{\rm P}$ is the number of datapoints, mainly because of the local density computation \citep{2019PaReL.128..551S}. 
Therefore, dealing with large datasets can be troublesome\footnote{In our experience, that is with $N_{\rm P} \gtrsim 10^5$.}. 
To work around this problem, we decided to transpose the algorithm to a grid-based approach. 
This allows to considerably reduce the number of datapoints and to filter out noisy points in sparse areas \citep[see, e.g.,][]{2018IEEE...84.513X}. 
Moreover, our initial dataset is already grid-based, since we use velocity fields from numerical simulations or observations. 
Therefore it is natural to force the EVCs and the resulting clusters to be defined on the same grid. 

Another important point is that we seek to identify vortices, which can have two orientations with respect to an axis, clockwise or counter-clockwise. However, the CFSFDP method evaluates all datapoints equally and does not consider the orientation of the local curvature they stem from. This can lead to an incorrect single identification when two vortices with opposite orientation stand close to each other. To avoid this situation, we classify EVCs as clockwise or counter-clockwise based on the orientation of the flow curvature. 
This can be easily achieved by checking the sign of the swirling strength $\lambda$, as done in Sect.\,\ref{subsubsec:curvature_radius}. In this way, EVCs belonging to each one of the two classes are separately clustered.   

The grid-vortex adaptation of the CFSFDP algorithm proceeds as follows.
The EVCs coordinates are rounded to coincide with the grid of the original velocity field. Then we count how many EVCs there are in each grid cell, differentiating between clockwise and counter-clockwise EVCs. Every EVC associated with a clockwise curvature counts as $-1$, while counter-clockwise EVCs count as $+1$.
The resulting sum for each grid cell is stored in the grid cardinality $s$. In this way, large amounts of clockwise EVCs rounded to the same grid cell generate a high negative cardinality $s$ value, while a high positive value of $s$ corresponds to a cluster of counter-clockwise EVCs.
We ignore all grid cells that have $|s| \leq 1$, since they do not contain any EVC or the one they contain can be considered as noise. From here on, all grid cells having $|s| > 1$ are labeled as Grid-Estimated Vortex Centers (G-EVCs) and their collection forms the G-EVC map.

The next step is to compute the local density $\rho$ of G-EVCs instead of single EVCs. 
To do so, we modify Eq.\,(\ref{eq:def_density}) to take into account the grid cardinality value and the sign of each G-EVC, which for a generic point $i$ is,
\begin{align}
    \rho_i^{\rm +} = |s_i| + \sum_{j,\, s_j > 0} |s_j|\, \chi ( d_{ij},\, d_c )\,, \nonumber\\
    \rho_i^{\rm -} = |s_i| + \sum_{j,\, s_j < 0} |s_j|\, \chi ( d_{ij},\, d_c )\,,
    \label{eq:density_grid}
\end{align}
where $\rho^{\rm +}_i$ and  $\rho^{\rm -}_i$ are the local densities of G-EVCs with positive (counter-clockwise) and negative (clockwise) cardinality, respectively. For the kernel $\chi$ one can use either Eq.\,(\ref{eq:density_kernel_1}) or Eq.\,(\ref{eq:density_kernel_2}). The first term, that is $|s_i|$, accounts for the number of EVCs contained in each G-EVC $i$, while the sum accounts for the local density due to neighboring G-EVCs. 
For the $\delta$ criterion, we use Eq.\,(\ref{eq:distance}) but limiting the calculation to the G-EVCs having the cardinality of the same sign. Hence, as for the density, we get two spacing criteria $\delta^+_i$ and $\delta^-_i$ which refer to the two classes (orientations) of G-EVCs. 

This procedure greatly reduces the required computational cost while keeping a high level of precision. 
To speed-up the calculation even more or improve the accuracy, the grid can in principle be respectively coarsened or refined at will. Moreover, the clustering can be performed independently for clockwise and counter-clockwise vortices, therefore this process can be parallelized.

Figure \ref{fig:demo_doublevortex} demonstrates this procedure for an artificial flow composed of two Lamb-Oseen vortices of opposite orientation.
Moreover, we added a random velocity field smoothed with a Gaussian filter, which serves as noise.
The two vortices are defined through Eq.\,(\ref{eq:lamb_oseen}). In more detail, the clockwise vortex is smaller in size but stronger in rotational speed with Lamb-Oseen parameters $\alpha = 1.256$, $r_{\rm max} = 0.07$, and $v_{\rm max} = -0.9$, while the counter-clockwise has $\alpha = 1.256$, $r_{\rm max} = 0.2$, and $v_{\rm max} = 0.6$, hence it is larger but slower. 
The average strength of the noise velocity field is $|v_{\rm noise}| = 0.226$.
The Cartesian grid has $200\,\times\,200$ grid points and covers an arbitrary domain of size $1.0\times1.0$.

The Rortex criterion captures the curvatures in the flow induced by the two vortices, as demonstrated in the top panel by the two large or intensive patches of positive (green) and negative (pink) $R$, respectively. The random motions caused by the Gaussian noise induce spurious signals ($R \neq 0$) in the Rortex map.
These signals are not related to the vortices, but to perturbations in the velocity field which cause the presence of local curvature in the flow. However, the flow in these places does not form a full circular or spiral pattern, and therefore these signals should not be regarded as vortices. 
If every connected region where $R \neq 0$ is identified as a vortex structure, then all these spurious signals erroneously represent different vortical regions. Hence, the necessity to consider the global properties of the flow in a more elaborated identification algorithm.

The bottom panel of Fig.\,\ref{fig:demo_doublevortex} shows the grid cardinality $s$ of the G-EVC map. 
We notice how most of the EVCs cluster near the centers of the two vortices, resulting in G-EVCs with high positive values of $s$ near the center of the counter-clockwise vortex and high negative values around the clockwise one. 
The estimation is not as precise as the one shown in Fig.\,\ref{fig:demo_lamboseen}, as many points are scattered in the surroundings of the vortex centers. 
This is due to the presence of noise and to the interaction between the flows of the two vortices. 
Nevertheless, the peaks of $|s|$ clearly indicate the presence of two vortex centers.
The location of these centers correspond to the approximate location of the original centers of the two vortex models, which are indicated by two blue circles in Fig.\,\ref{fig:demo_doublevortex}. 

In addition there are small patches of G-EVCs scattered in the map. 
These points show uniformly low values of $s$, which is precisely the behavior expected for incoherent local curvature perturbations in the flow. Random curvatures in the flow show up in the mathematical criteria for the identification of vortices, as seen in the top panel of Fig.\,\ref{fig:demo_doublevortex}, but will not be selected as cluster centers since they are characterized by a relatively low density $\rho^{\pm}_i$. 
The decision process is subject of the following subsection.


\subsubsection{Decision}
\label{subsubsec:decision}

\begin{figure}
	\centering
	\resizebox{\hsize}{!}{\includegraphics{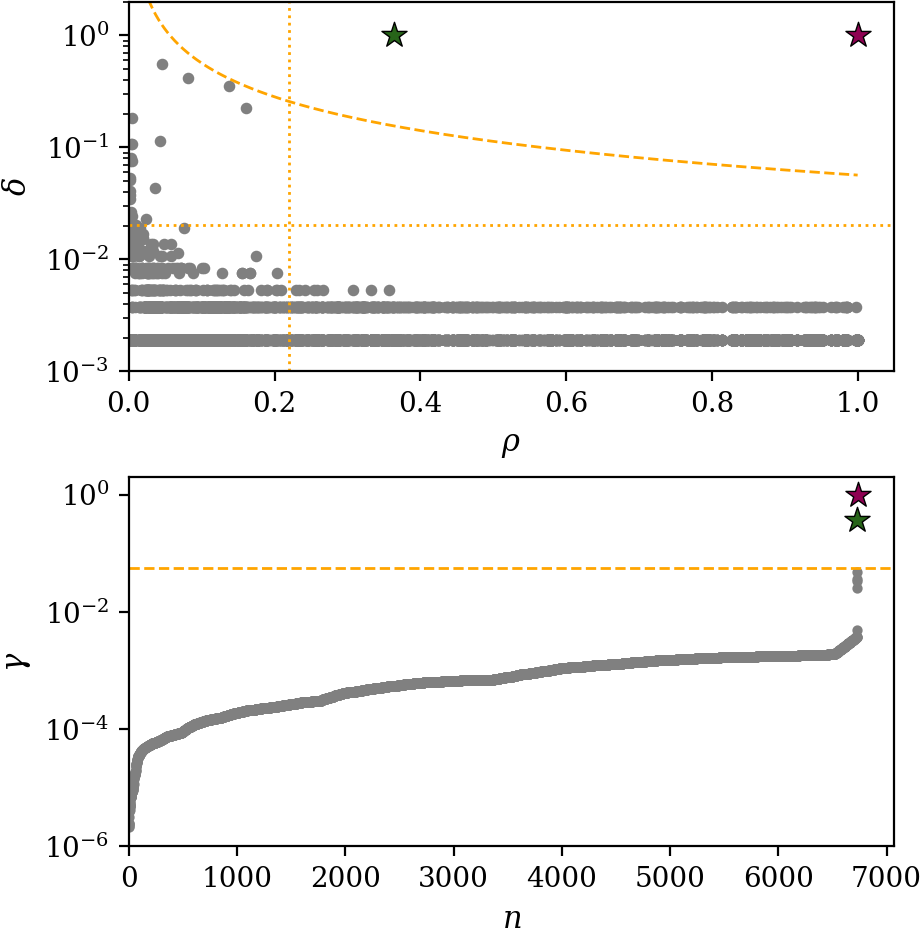}}
	\caption{Decision graphs for the artificial Lamb-Oseen vortex flow of Fig.\,\ref{fig:demo_doublevortex} based on the $\rho-\delta$ thresholds of Eq.\,(\ref{eq:rho_delta_thresholds}) (\textit{top}) and on the $\gamma$ threshold of Eq.\,(\ref{eq:gamma_threshold}) (\textit{bottom}). 
	In the bottom panel, $n$ represents the datapoint number in increasing order of $\gamma$.
	Dotted lines represent the $\rho$ and $\delta$ thresholds, while dashed lines show the $\gamma$ threshold. Each point corresponds to a EVC. The green and pink starred EVCs satisfy both thresholds and are therefore selected as cluster centers.}
	\label{fig:demo_doublevortex_decision}
\end{figure}

The next step is to select the cluster centers based on the local density $\rho = [\rho^{+},\rho^{-}]$ and the spacing $\delta = [\delta^+, \delta^-]$ quantities. A universal selection criteria is still under debate. A common strategy is to define a couple of thresholds, $\rho_{\rm th}$ and $\delta_{\rm th}$, based on the distribution of $\rho$ and $\delta$ within the dataset \citep[][]{2014Sci...344.1492R}. Every point having local density and spacing larger than both thresholds is selected as a cluster center. However, the choice of the thresholds is arbitrary and depends on the type of dataset.
For example, the local density and spacing thresholds can be defined as,
\begin{equation}
    \rho_{\rm th} = p_{\rho} \mu_{\rho}~,~~ \delta_{\rm th} = p_{\delta} \sigma_{\delta}\,, \label{eq:rho_delta_thresholds}
\end{equation}
where $\mu_{\rho}$ is the mean value of the local density $\rho$, $\sigma_{\delta}$ is the standard deviation of the spacing $\delta$, and $p_{\rho},\,p_{\delta} > 0$ are free parameters. Moreover, to avoid biased thresholds $\rho_{\rm th}$ and $\delta_{\rm th}$ when evaluations Eq.\,(\ref{eq:rho_delta_thresholds}) in the grid-adapted version of the algorithm, one also needs to take the number of datapoints (EVCs) hidden in each G-EVC into account. Therefore, one should add $s_i - 1$ mock datapoints with $\rho = \rho_i$ and $\delta = \frac{1}{2} \Delta x$ values for each G-EVC $i$, where $\Delta x$ is the grid cell size. In this way, each G-EVC $i$ will be represented by one datapoint with $\rho = \rho_i$ and $\delta = \delta_i$ in the cluster center decision process, while the mock datapoints with $\rho = \rho_i$ and $\delta = \frac{1}{2} \Delta x$ account for the remaining EVCs which are considered as close neighbors. 

Alternatively, one can set a threshold for the $\gamma$ criterion, where $\gamma = \rho \delta $. This quantity naturally highlights datapoints having both local density and spacing relatively high. The choice of the threshold is once more arbitrary, and we opt for the following formulation,
\begin{equation}
    \gamma_{\rm th} = p_{\gamma} \delta_{\rm min} \rho_{\rm max}\,, \label{eq:gamma_threshold} 
\end{equation}
where $p_{\gamma} > 0$ is again a free parameter, while $\delta_{\rm min}$ and $\rho_{\rm max}$ are the minimum and the maximum of the spacing $\delta$ and local density $\rho$ distributions, respectively. This choice ensures that, for $p_{\gamma} > 1$, datapoints with large local density but very small distance (that is close-neighbors) are not selected as cluster centers. Moreover, it relies on only one free parameter.
Another possibility is to sort the values of $\gamma$ in increasing order and identify a knee point in the curve, as proposed by \citet{2016CompComm...10.1109W}, 
but we have not tested this option so far.

Figure \ref{fig:demo_doublevortex_decision} shows two commonly used decision graphs for the visualization of the cluster centers selection \citep{2014Sci...344.1492R}. 
The top panel shows the normalized spacing criterion $\delta_i/\delta_{\rm max}$ and the normalized density $\rho_i/\rho_{\rm max}$ for each datapoint, while in the bottom panel the $\gamma$ criterion is plotted for each datapoint in increasing order of $\gamma$.
The thresholds in $\rho$, $\delta$, and $\gamma$, defined through Eqs.\,(\ref{eq:rho_delta_thresholds}) and (\ref{eq:gamma_threshold}), are indicated as dotted and dashed orange  lines, respectively. The chosen parameters are $p_{\rho} = 1.0$, $p_{\delta} = 0.5$, and $p_{\gamma} = 15.0$. There are two datapoints, marked with a star, that satisfy both selection criteria and are therefore identified as cluster centers. 


\subsubsection{Vortex identification}
\label{subsubsec:vortex_identification}
\begin{figure}
	\centering
	\resizebox{\hsize}{!}{\includegraphics{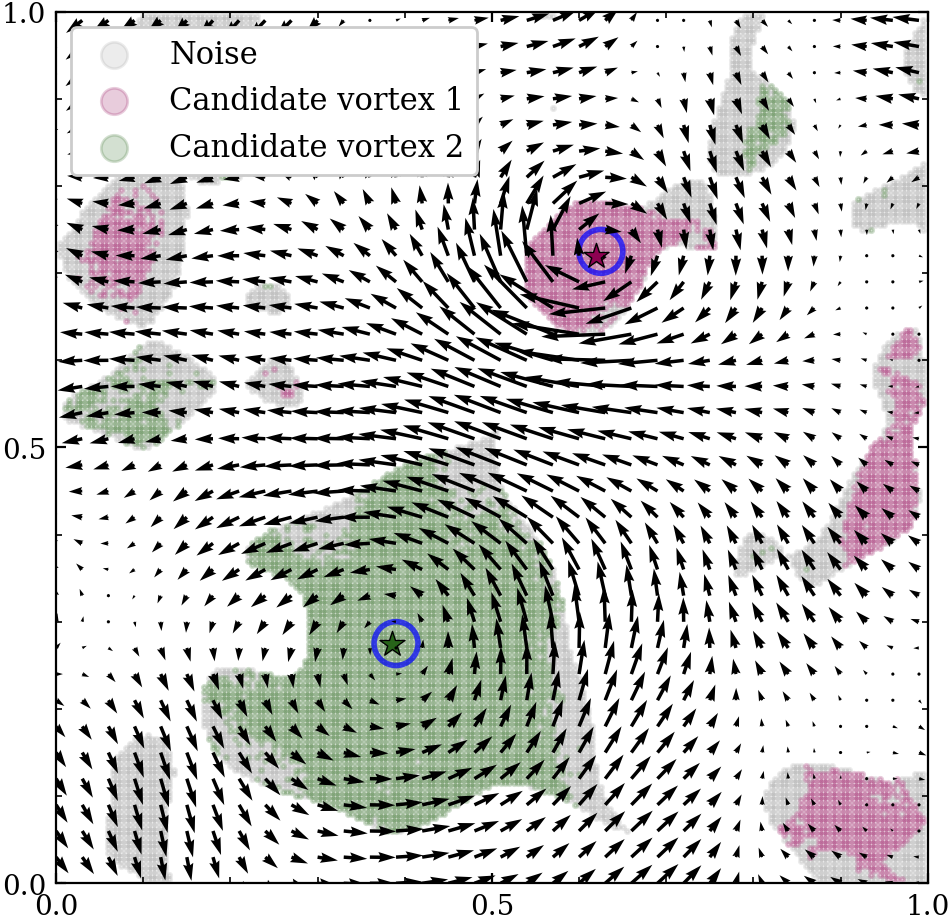}}
	\caption{Groups of 
	grid cells
	of candidate vortices (green and pink pixels) obtained by applying the grid-adapted CFSFDP algorithm to the G-EVC map of the artificial flow shown in Fig.\,\ref{fig:demo_doublevortex}. The cluster centers are marked with a star, while the blue circles show the approximate location of the two Lamb-Oseen vortex centers. Noisy grid cells are gray shaded.
	}
	\label{fig:demo_doublevortex_nonoise}
\end{figure}
\begin{figure}
	\centering
	\resizebox{\hsize}{!}{\includegraphics{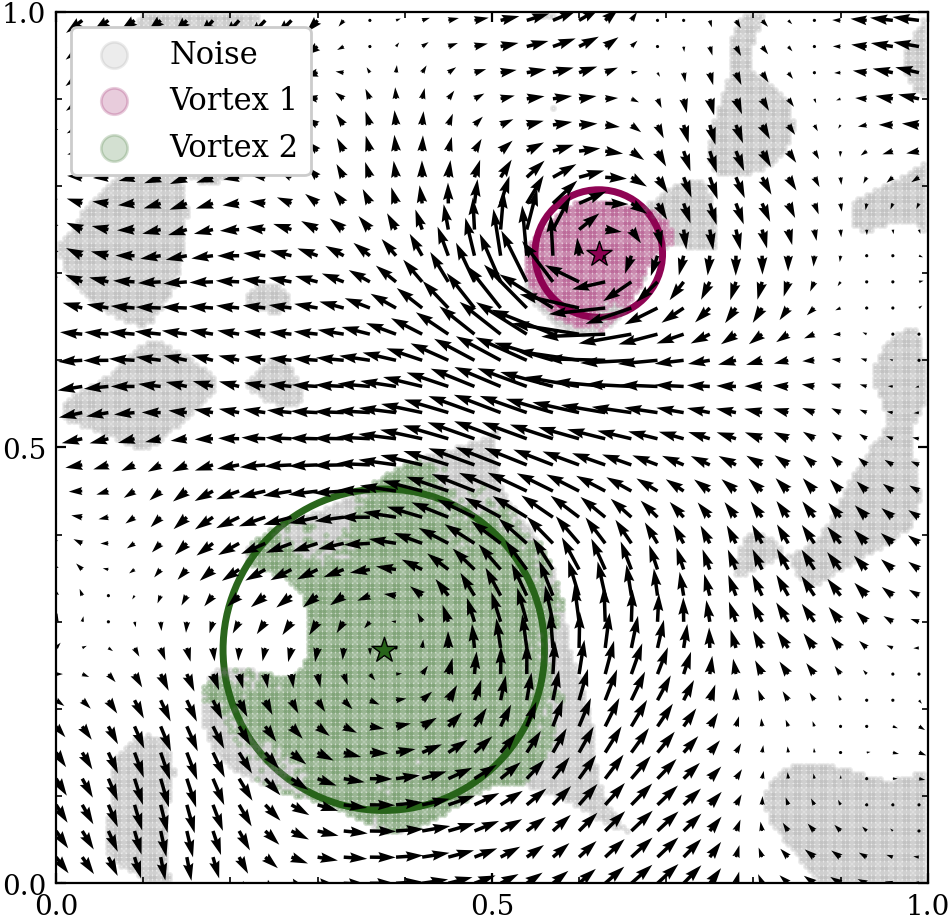}}
	\caption{Vortices identified in the artificial flow shown in Fig.\,\ref{fig:demo_doublevortex}. The vortex centers are marked with a star, while the circles show their effective size according to Eq.\,(\ref{eq:cluster_radius}). 
    Grid cells
	determined to be noise are gray shaded.}
	\label{fig:demo_doublevortex_result}
\end{figure}

For each cluster center that fulfills the decision criteria set by the thresholds Eq.\,(\ref{eq:rho_delta_thresholds}) or (\ref{eq:gamma_threshold}), we proceed with the formation of an associated cluster of datapoints (in our case, G-EVCs). For that, we assign each one of the remaining datapoints, not already identified as a cluster center, to the same cluster of its nearest neighbor of highest density \citep{2014Sci...344.1492R}. 
The single EVCs are assigned to the same cluster of the G-EVCs they belong to.

At the end of this process, every G-EVC and consequently every EVC will belong to a cluster.
Since there is a one-to-one correspondence between the EVC points and the grid cells they derive from, each cluster of EVCs corresponds to a group of grid cells showing local curvature in the velocity field.
Therefore, each group of grid cells corresponds to a vortex candidate with its center given by the associated G-EVC cluster center. 
Figure \ref{fig:demo_doublevortex_nonoise} shows the two groups of grid cells
(vortex candidates) found for the artificial flow of Fig.\,\ref{fig:demo_doublevortex}. The cluster centers found in Fig.\,\ref{fig:demo_doublevortex_decision} represent the centers of the identified candidate vortices and are marked with stars. The two main groups of grid cells belonging to the Lamb-Oseen vortical structures are identified by the method. 

\subsubsection{Noise removal}
\label{subsubsec:noise_removal}
In Fig.\,\ref{fig:demo_doublevortex_nonoise} are also shown patches of grid cells already identified as noise. These are grid cells characterized by $R\neq0$, as shown in Fig.\,\ref{fig:demo_doublevortex}, that however have already been discarded because their EVCs are found either outside of the computational domain or alone in a grid cell of the G-EVC map. Therefore, we can assume that the curvature of the flow identified in these cells is not part of a coherent vortical structure. The associated EVCs are not considered in the clustering process.  

There are also a few isolated patches of grid cells that have been associated to one of the candidate vortices by the clustering algorithm although they are clearly not related to them. These grid cells have not been discarded yet because their EVCs are accidentally associated with one of the two main clusters, or because they are part of other local clusters which however do not satisfy the decision criteria of the clustering algorithm. Indeed, a few more groups of G-EVCs with low grid cardinality $|s| \lesssim 20$ are visible in the bottom panel of Fig.\,\ref{fig:demo_doublevortex}. Since these EVCs are considered in the cluster identification process, they are also inevitably associated to a cluster. Therefore, the corresponding grid cells belong to one of the candidate vortices. We seek to label these grid cells as noise as well and remove them from the corresponding candidate vortex.

In order to do so, we consider a grid cell that has been associated to a candidate vortex as noise if either its position or its EVC is ``sufficiently distant'' from the center of the cluster they belong to. In this way, grid cells can be removed if they or their EVCs are spatially unrelated to the vortex structure they supposedly belong to.
We define ``sufficiently distant'' a distance relative to the effective radius of the candidate vortex, $r_{\rm eff}$, which is estimated based on the area covered by the grid cells belonging to that structure as,
\begin{equation}
    r_{\rm eff} = \sqrt{\frac{N_{\rm c}}{\pi}} \Delta x \,, \label{eq:cluster_radius}
\end{equation}
where $N_{\rm c}$ is the number of grid cells that form the candidate vortex and $\Delta x$ is the grid cell size. Hence, grid cells for which the distance between the cluster center and their position or their corresponding EVC location are larger than $p_{\rm r} r_{\rm eff}$, where $p_{\rm r}$ is a free parameter, are labeled as noise and removed from the candidate vortex structure. We note that this is an iterative process because the removal of grid cells changes $N_{\rm c}$ and with it the criterion $p_{\rm r} r_{\rm eff}$.

Finally, we try to identify and remove candidate vortices that do not show full circular patterns in their flow. 
For example, a sharp deviation in the flow induces some degree of local curvature ($R \neq 0$) and possibly a cluster of EVCs, but it should not be classified as a vortex because the trajectories of particles in such a flow do not spiral. To differentiate non-spiraling coherent curvatures in the flow 
from actual vortices, we use the fact that grid cells belonging to a perfect vortex are isotropically distributed around the cluster center, while for non-spiraling coherent curvatures a strongly asymmetric distribution can be expected. Therefore, the effective radius computed with Eq.\,(\ref{eq:cluster_radius}) results in a good estimate of the size of a vortical structure only, while for non-spiraling coherent curvatures it underestimates their radius. Repeating the procedure for removing distant grid cells while updating with each iteration the number of grid cells $N_{\rm c}$ and the effective radius $r_{\rm eff}$, for a parameter $p_{\rm r}$ sufficiently close to $1$, a cluster stemming from a non-spiraling coherent curvature in the flow should be depleted of all grid cells composing it, while a vortex should maintain approximately its total number of grid cells and size. 

Figure \ref{fig:demo_doublevortex_result} shows the final result, where the outlying patches of grid cells have been classified as noise with the procedure just described, while the two Lamb-Oseen vortices have been successfully identified. The effective radii of the two vortices, computed using Eq.\,(\ref{eq:cluster_radius}), are $0.19$ for the anti-clockwise vortex and $0.08$ for the other one. Considering the interaction between the two vortices and the perturbations due to the random velocity field, the estimated radii are very close to the initial values, which are $0.20$ and $0.07$ respectively. 
The location of the vortex centers cannot be determined precisely because of the random Gaussian velocity field that perturbs the flow and therefore shifts, although slightly only, their original position. Nevertheless, the identified vortex centers are found in the vicinity of the original positions of the Lamb-Oseen vortices, marked by blue circles in Figure \ref{fig:demo_doublevortex_nonoise}, which confirms the functioning of the method.

We implemented the new vortex identification method and the automated algorithm in a Python package named "SWirl Identification by Rotation-centers Localization" (SWIRL). The code is open source and can be found online\footnote{\url{https://github.com/jcanivete/swirl}}. In Appendix \ref{app:algorithm} we review the structure of the code with a step-by-step flowchart.
%
%

\section{Application and discussion}
\label{sec:application_and_discussion}
In this section, we test the vortex identification algorithm on more complex vortical flows than those of Sect.\,(\ref{sect:method}). First, we apply the SWIRL code to an artificial flow composed of nine vortices of different sizes and strengths and a Gaussian noisy signal. Then, in order to test the reliability of the code on a more turbulent flow, we apply it to a two-dimensional simulation of a magneto-hydrodynamical (MHD) Orszag-Tang vortex.

%
%

\subsection{Artificial vortex flow}
\label{subsec:application_LO}
\begin{figure}
	\centering
	\resizebox{\hsize}{!}{\includegraphics{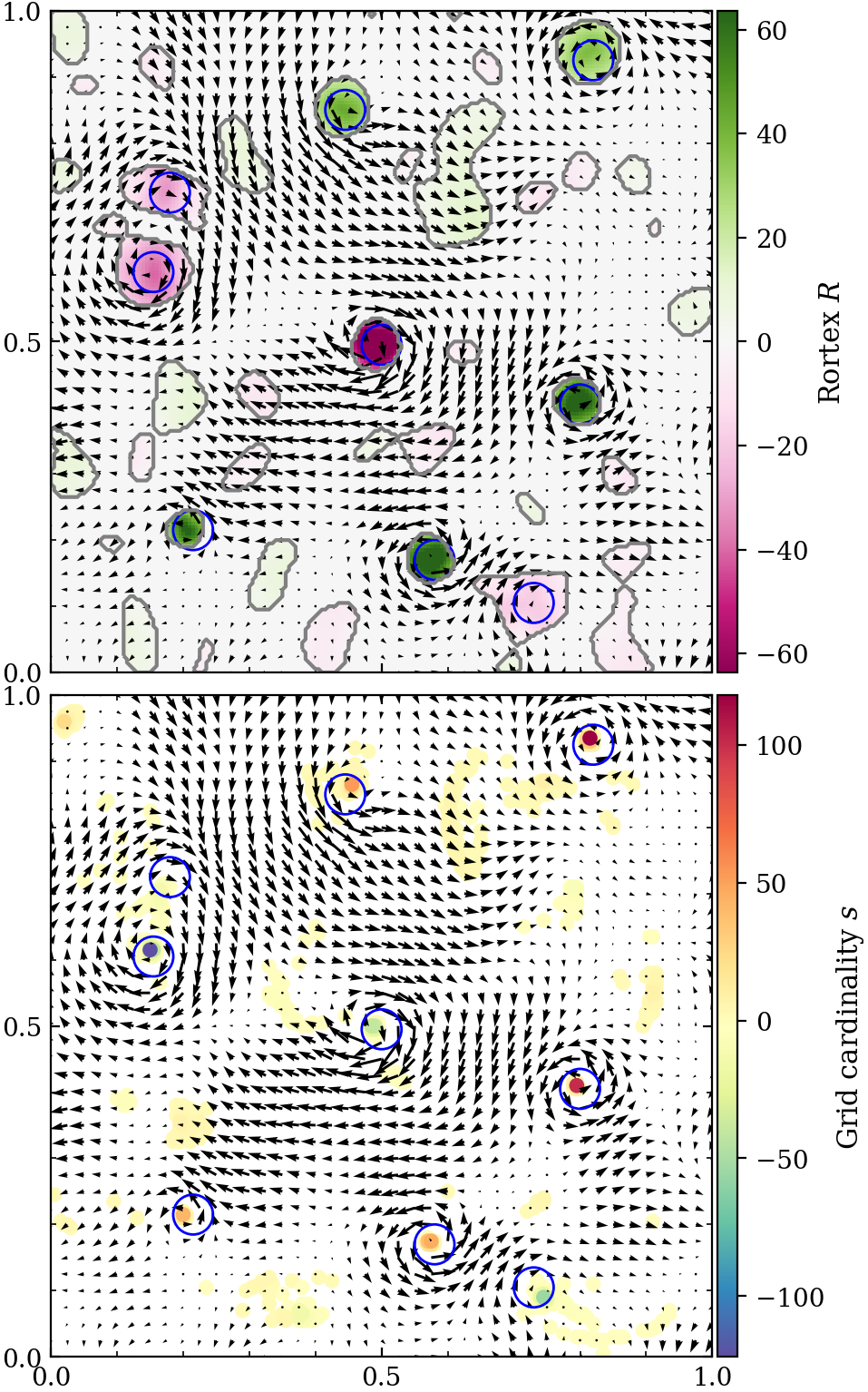}}
	\caption{Artificial flow with nine different Lamb-Oseen vortices superposed by a Gaussian smoothed random velocity field. \textit{Top panel}: Rortex map. \textit{Bottom panel}: G-EVC map. Blue circles denote the position of the Lamb-Oseen vortex centers. The velocity field is represented by a vector field in both panels.
	The regions where $R \neq 0$ are outlined with a gray contour in the Rortex map.
	}
	\label{fig:test_multiplevortex}
\end{figure}
\begin{figure}
	\centering
	\resizebox{\hsize}{!}{\includegraphics{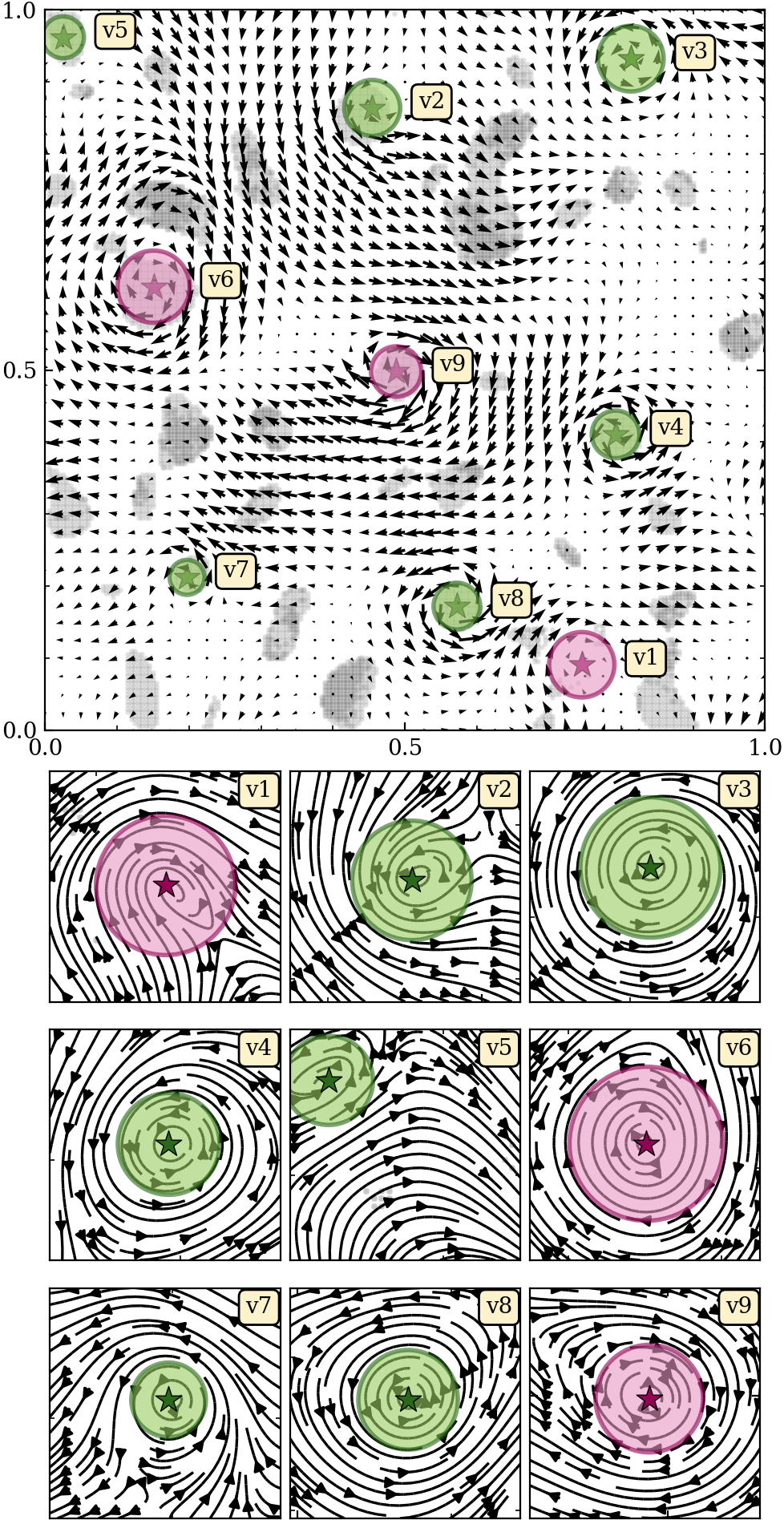}}
	\caption{Identified vortices in the artificial flow of Fig.\,\ref{fig:test_multiplevortex}. \textit{Top panel}: Position and the size of nine identified vortices. Green colored vortices rotate counter-clockwise, the pink ones rotate clockwise. The velocity field is displayed by a vector plot. 
    Grid cells
	labeled as noise are shown in gray. \textit{Bottom panels}: Close-ups on the different vortices. The velocity field is represented by streamlines. Stars denote the center of the vortices and the radii are computed according to Eq.\,(\ref{eq:cluster_radius}).}
	\label{fig:test_multiplevortex_result}
\end{figure}
The artificial flow is composed of nine random Lamb-Oseen vortices and a Gaussian smoothed noisy velocity field as shown in Fig.\,\ref{fig:test_multiplevortex}. The parameters $r_{\rm max}$, $v_{\rm max}$, and $\alpha$ of the Lamb-Oseen model, as well as the coordinates of the vortex centers, are randomly generated. The size of the grid is $200\times 200$ points covering a domain of $1.0\times 1.0$ in arbitrary units. In the top panel, we show the Rortex map derived from the composed velocity field. After the computation of the EVC map, we obtain the grid cardinality $s$ map shown in the bottom panel of the same figure. The approximate coordinates of the Lamb-Oseen vortex centers are marked with blue circles in both panels. We notice peaks of positive and negative grid cardinality $s$ in correspondence to counter-clockwise and clockwise curved flows, respectively. We then proceed to an automated identification of vortices with the method presented in Sect.\,\ref{sect:method} and implemented in the SWIRL code.

The result is shown in the top panel of Fig.\,\ref{fig:test_multiplevortex_result}. 
For simplicity,  we show the identified vortices with a colored disk representing their effective circular area, instead of showing the individual grid cells that compose them. The size of the disk is estimated with Eq.\,(\ref{eq:cluster_radius}),
while the color indicates their orientation. Grid cells that have been identified as noise are instead colored in gray. Our algorithm identifies nine vortices in total, but only eight of those are related to the artificial Lamb-Oseen vortices inserted in the velocity field. The vortex placed at coordinates $(x,y) \sim (0.20, 0.75)$ has been discarded and recognized as noise. Inspecting the velocity field around that location with the help of the vector plot reveals that the flow there performs a non-spiraling coherent curvature. 
This is due to the presence of a nearby stronger vortex at coordinates $(x,y) \sim (0.20, 0.60)$, identified by the algorithm and labeled as v6. Since both vortices have the same orientation, the velocity field is dominated by the stronger one (vortex v6) and the flow is distorted around the weaker one.

The algorithm however identifies a vortex, labeled v5, around $(x,y) \sim (0.05, 0.95)$, that is not associated with any of the artificially generated vortices, but stems from the background random velocity field. The bottom panels of Fig.\,\ref{fig:test_multiplevortex_result} show close-ups on the nine detected vortices with instantaneous streamlines of the velocity field instead of vector plots. The identified vortices are characterized by bounded streamlines around the vortex centers, which are marked by colored stars.

The identification of vortex v5 demonstrates an important feature of our algorithm. From the vector plot of Fig.\,\ref{fig:test_multiplevortex}, this vortex is not easily recognizable from visual inspection because the velocity field in that region is weak compared to the rest of the domain.
Moreover, the values of the Rortex criterion are relatively small because of the weak angular velocity, as we can see from Fig.\,\ref{fig:test_multiplevortex}. Traditional algorithms based on mathematical criteria might miss this vortex, since for those a threshold in the criteria is likely put in place to filter out noise. In that case, the signal due to the vortex would be removed. 
However, the method presented in this paper does not require a threshold on the Rortex criterion, hence it can also detect weak vortical flows.      
%
%

\subsection{Orszag-Tang vortex test}
\label{subsec:application_OT}

\begin{figure}
	\centering
	\resizebox{\hsize}{!}{\includegraphics{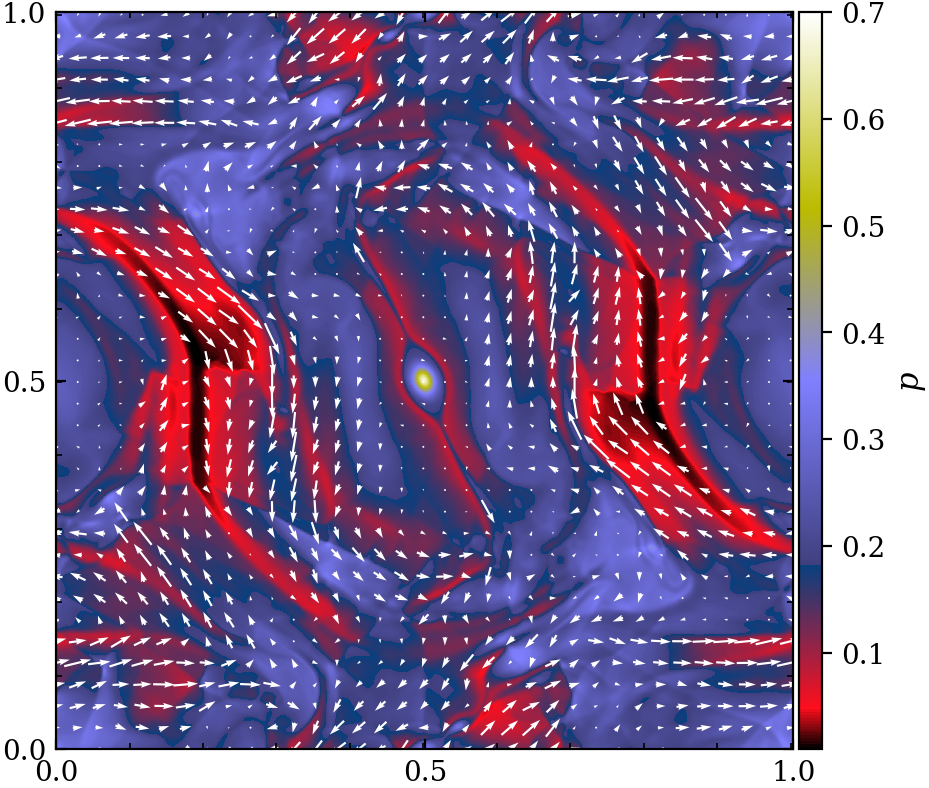}}
	\caption{Pressure $p$ map of the two-dimensional MHD Orszag-Tang test at $t=1.0$ carried out with the CO5BOLD code. The velocity field is represented by white arrows.}
	\label{fig:test_ot_pressure}
\end{figure}
\begin{figure}
	\centering
	\resizebox{\hsize}{!}{\includegraphics{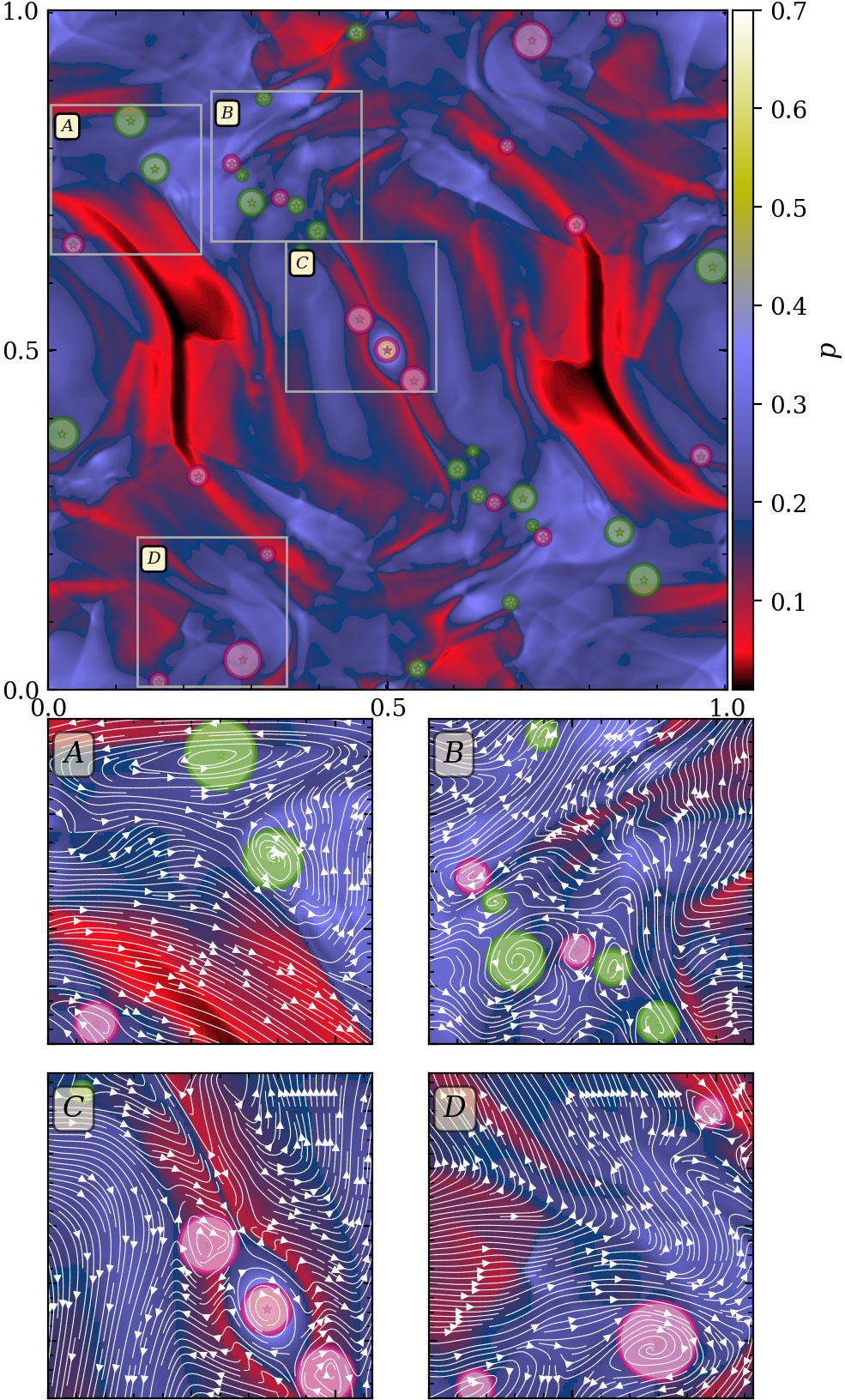}}
	\caption{Identified vortices in the Orszag-Tang test flow at $t=1.0$. \textit{Top panel}: Position and the size of the identified vortices. Green colored vortices rotate counter-clockwise, while the pink ones rotate clockwise. \textit{Bottom panels}: Close-ups on four different regions of the flow with identified vortices. The velocity field is represented by instantaneous streamlines. The size of each vortex is computed according to Eq.\,(\ref{eq:cluster_radius}).
	}
	\label{fig:test_ot_swirls}
\end{figure}

The Orszag-Tang vortex problem \citep[][]{1979JFM....90..129O} is one of the most widely known benchmark tests for MHD numerical schemes. It is a two-dimensional problem and the initial conditions on a domain of size $1.0\,\times\,1.0$ are given as follows,
\begin{align}
    \boldsymbol{v} & =  \Big( -\sin{(2\pi y)},\, \sin{(2\pi x)}\Big)\,, \nonumber \\
    \boldsymbol{B} & =  \Big( - B_0 \sin{(2\pi y)},\, B_0 \sin{(4\pi x)}\Big)\,, \nonumber \\
    \rho & =  \gamma p_0\,, \label{eq:OT_IC}
\end{align}
where $B_0 = (4\pi)^{-1/2}$, $\gamma = 5/3$, and $p_0 = 5/(12\pi)$. The initialized flow is unstable and breaks down into turbulence very quickly. We ran the Orszag-Tang problem with the CO5BOLD code \citep[][]{2012JCoPh.231..919F} 
on a $512\,\times\,512$ grid. The pressure $p$ and the large scale velocity field at $t=1.0$ are shown in Fig.\,\ref{fig:test_ot_pressure}. 

Multiple vortices are expected to form in the simulated turbulent flow. Given the complexity of the dynamics and the influence of magnetic fields, these vortices are most likely different from the Lamb-Oseen models we used in Sect.\,\ref{subsec:application_LO}. Nevertheless, our approach is not limited to Lamb-Oseen vortex models since the Rortex criterion measures, by definition, the rigid-body rotational component of any velocity field. Therefore, we expect our methodology to be successful also in more complex flows, such as the Orszag-Tang vortex test or realistic simulations of the solar atmosphere. For example, \citet[][]{2020ApJ...898..137S} showed that the profiles of the tangential velocity of vortices identified in numerical simulations of the solar atmosphere can be well fitted by a cubic polynomial. 

We applied the SWIRL code to the simulated velocity field at $t=1.0$ and the results are shown in Fig.\,\ref{fig:test_ot_swirls}. To enhance the robustness of the identification process, we employed the ``multiple stencil'' approach (see Sect.\,\ref{subsubsec:curvature_radius}) with stencils of $1$, $3$, $5$, and $7$ grid cells. In total, 37 swirls are identified in a mirror-like disposition, which evidences the symmetry of the problem. In the bottom panels of the same figure we can appreciate the turbulent nature of the flow as shown by the instantaneous streamlines. Almost all the identified vortices correspond to locations in the flow where the instantaneous streamlines form circular or spiral patterns. The only exception is the one at the top of the $B$ panel, where the flow seems to perform a sharp U turn but not a full vortical motion. 

It is important to notice that the streamlines shown in the bottom panels of Fig.\,\ref{fig:test_ot_swirls} do not reflect the trajectories that test particles would follow in such a flow, but stem from the instantaneous velocity field. Therefore, one must be cautions when investigating the morphology of velocity fields from a single time instance in particular when it changes on a timescale short compared to the rotational period of the vortical flow \citep[see, e.g.,][]{2013ApJ...776L...4S}. 
Nevertheless, we believe that the results demonstrate that the SWIRL code is able to correctly identify most of the vortical motions present in the simulated flow, even in the presence of turbulence and magnetic fields. Applications to solar physical flows are planned to be presented in a follow-up paper.

%
%

\section{Summary and conclusions}
\label{sec:conclusions}
In this paper, we presented a new method and an automated algorithm for the identification of vortical motions based on the velocity field alone. The core computation of this method is an estimation of curvature centers for all points in the velocity map exhibiting some degree of local curvature. For that, we combine an innovative and highly reliable mathematical criterion, the Rortex, with the global information of the flow derived from the velocity field.
Hence, the method can be considered a hybrid between classical methods based on mathematical criteria and morphological methods. 
To our knowledge, this is the first time that such type of a vortex detection algorithm has been suggested, both in the domain of astrophysics and of computational fluid dynamics.
We classify the estimated centers of rotation, or EVCs, with a grid-adapted version of the modern clustering algorithm CFSFDP. This automated procedure can piece together all the points that share a common center of rotation and therefore form a vortex. A first implementation of the algorithm, called SWIRL, is open source and can be found online.   

We tested our code on a non-trivial artificial velocity field composed of nine different Lamb-Oseen vortices and a background Gaussian noise. The algorithm correctly detected all vortical motions of different angular velocities and sizes present in the flow, even one that was very weak and was generated by the random perturbations of the background noise. 
We also tested the SWIRL code with a MHD numerical simulation of a Orszag-Tang vortex system. The simulated flow is turbulent in nature and influenced by the magnetic field, hence the identification process is much more complex than in the artificial case and misidentifications can be expected. Nevertheless, the results demonstrate the high level of accuracy of our methodology.

Therefore, the SWIRL code is a reliable tool for the study of vortical motions in complex and magneto-hydrodynamical astrophysical systems. It is robust against noise, turbulence, and shear flows; at the same time, weakly rotating vortices do not pose problems as thresholds in the Rortex values are not necessary. The vortex center is naturally given by the EVCs cluster center, while proper boundaries can be defined through the collection of grid cells composing the vortex structure in accordance with the definition of \citet[][]{1979rdte.book..309L} that a vortex is ``a multitude of material particles (grid cells) rotating around a common center''. In this way, one can determine physical quantities of interest in any point of the vortical region, allowing for a trustworthy analysis on the vortex properties and dynamics. Finally, the SWIRL code can be directly applied to the velocity field of numerical simulations as well as to previously derived velocity fields of observations.

Regarding computational performance, the method is very fast on small to medium grid ranges. 
For example, the full identification process carried out on a $200\times200$ grid for the artificial flow of Sect.\,\ref{subsec:application_LO} took $\sim1\,{\rm s}$ on a single CPU. For the Orszag-Tang test instead, given the complexity of the flow and the larger size of the grid $(512\,\times\,512)$, the computation took $\sim 60\,{\rm s}$. It can become relatively slow on large domains, especially when numerous vortices are present, the main cause being the clustering step. 
However, one can always adopt a ``divide and conquer'' approach and run the code on smaller portions of the grid separately and combine the results in the end. In principle, it is also possible to parallelize the process. 
Finally, it is worth to mention that no cut-off value for the Rortex criterion needs to be imposed as done in many methods based on mathematical criteria,
because EVCs due to noisy signals are scattered randomly in the domain and do not form clusters. 

There are, however, two drawbacks to be noted. First, the method is not strictly Galilean invariant since the vortex structure is required to be at rest with respect to translations for the accurate estimation of its center of rotation. For solar applications this is not much of a concern since vortices are usually anchored within intergranular lanes or vertices of intergranular lanes and move slowly relative to the vortical flow speed \citep[see, e.g.,][]{2022...ISSI}. 
However, it could be a matter of concern in more dynamical scenarios. Second, the parameters of the clustering algorithm call for some fine tuning to perform as expected. 

The algorithm can be further developed and improved. First of all, our implementation is limited to two dimensions. In principle, the computation of a fully three-dimensional EVC map should be straightforward,
as Eqs.\,(\ref{eq:estimated_direction}) and (\ref{eq:estimated_radius}) are valid in three-dimensions as well. In that case, one would obtain a three-dimensional distribution of EVC points. However, the clustering process is likely to be computationally extremely expensive, because of the extra dimension and the consequently larger datasets.
A thorough study on multi-dimensional clustering algorithms is therefore required for a possible development in this direction. On the other hand, an improvement of the current status of the algorithm is also possible. The grid-based implementation of the CFSFDP clustering algorithm can be further refined to make it more accurate and robust. In particular, we would like to reduce the number of parameters required in order to make it as user-independent as possible. Techniques as the one proposed by \citet[][]{2016CompComm...10.1109W} 
could be implemented in the future to attain this goal.

Another area of possible improvement regards the accuracy of the computed EVCs. Indeed, Eqs.\,(\ref{eq:estimated_direction}) and (\ref{eq:estimated_radius}) are only approximate estimations of the radial direction and curvature radius, especially in complex and turbulent flows such as solar atmospheric ones. Improving the accuracy of these quantities, even slightly, would affect the clustering process and, in the end, the performance of our method.

In conclusion, the method presented in this paper represents a new and reliable technique for the detection and analysis of vortices in numerical simulations of turbulent and (magneto-)hydrodynamical flows. The publicly available implementation allows for a simple and quick usage of the algorithm. In a follow-up paper, we intend to apply our method to realistic numerical simulations of the solar and stellar atmospheres carried out with the CO5BOLD code.
In the future, we would also like to compare our method to other commonly used automated vortex identification algorithms and to draw rigorous statistics of vortex populations from high-resolution observations of the solar atmosphere. 
%
%

\begin{acknowledgements}
The authors acknowledge support by SNF under grant ID 200020\_182094. This work has profited from discussions with the team of K.\,Tziotiou and E.\,Scullion (conveners) ``The Nature and Physics of Vortex Flows in Solar Plasma'' and with the team of P. Keys (convener) ``WaLSA: Waves in the Lower Solar Atmosphere at High Resolution'' (\url{www.walsa.com}) 
at the International Space Science Institute (ISSI). 
We would like to sincerely thank the anonymous referee for the constructive comments and F.\,Riva (IRSOL) for providing us with the Orszag-Tang test model.
\end{acknowledgements}


%
%

\bibliographystyle{aa} 
\bibliography{biblio.bib} 

%
%

\begin{appendix}

\section{Analytical comparison between mathematical criteria}
\label{app:analytical_comparison}

In Sect.\,\ref{subsec:comparing_criteria} we show numerically that the Rortex is a mathematical criterion that accurately measures the rigid-body rotational part of 
the velocity field of a Lamb Oseen vortex model.
However, we can prove it analytically as well, and generalize this results to different kinds of vortical motions by considering a few examples.

We start with the simplest vortex model, namely the rotational vortex. Without loss of generality, we center the vortex flow in $(x,y) = (0,0)$, such that the associated velocity field can be defined as,
\begin{align}
    v_x & = -\alpha y\,, \nonumber\\
    v_y & = +\alpha x\,, \label{eq:rotational_vortex}    
\end{align}
where $\alpha>0$ is a constant that defines the strength and the orientation of the rotation. We can compute the vorticity, the swirling strength, and the Rortex criteria with Eqs.\,(\ref{eq:vorticity}), (\ref{eq:U_decomposition}), and (\ref{eq:Rortex}), respectively. The resulting values for these criteria are,
\begin{equation}
    \omega = \lambda = R = 2\alpha\,. \label{eq:rotational_vortex_result}
\end{equation}
Hence, for a simple rotational vortex, the three criteria give identical results. Moreover, the tangential component of the velocity is $v_{\theta} = \alpha r$, where $r = \sqrt{x^2 + y^2}$ is the radius in polar coordinates. Employing Eqs.\,(\ref{eq:period_criteria}) and (\ref{eq:period_analytical}) one can check that the three criteria yield the correct rotational period, that is $T = 2\pi/\alpha$. This is not a surprise since a rotational vortex behaves as a rotating rigid-body, and therefore consists of pure rotation. 

Next, we use a generalized version of the rotational vortex model where the tangential velocity scales as a power of the radius, that is,
\begin{equation}
    v_{\theta} = \alpha r^{\beta}\,, \label{eq:power_vortex}
\end{equation}
where $\beta \in \mathbb{R}$. The rotational vortex model is a special case of Eq.\,(\ref{eq:power_vortex}) with $\beta = 1$, while the irrotational vortex model can be obtained with $\beta = -1$. For the case of Eq.\,(\ref{eq:power_vortex}) the velocity field in Cartesian coordinates is given by,
\begin{align}
    v_x & = - \alpha r^{\beta - 1} y\,, \nonumber\\
    v_y & = + \alpha r^{\beta - 1} x\,. \label{eq:power_vortex_xy}
\end{align}

Using again Eqs.\,(\ref{eq:vorticity}), (\ref{eq:U_decomposition}), and (\ref{eq:Rortex}) one obtains the following values for the three criteria,
\begin{align}
    \omega & = \alpha r^{\beta-1} (1+\beta)\,, \nonumber \\
    \lambda & = 2\alpha r^{\beta-1} \sqrt{\beta}\,, \nonumber \\
    R & = 2 \alpha r^{\beta-1}\,. \label{eq:criteria_power_vortex}
\end{align}

In this case, each criterion yields a different value and therefore predicts a different period of rotation. The true period of rotation is given by Eq.\,(\ref{eq:period_analytical}) with $v_\theta$ given by Eq.\,(\ref{eq:power_vortex}), that is, $T = 2\pi/(\alpha r^{\beta - 1})$. Using Eq.\,(\ref{eq:period_criteria}) to estimate the period of rotation from the different criteria, we obtain,
\begin{align}
    T_{\omega} & = \frac{4\pi}{\alpha r^{\beta-1} (1+\beta)}\,, \nonumber \\
    T_{\lambda} & = \frac{4\pi}{\alpha r^{\beta-1} \sqrt{\beta}}\,, \nonumber \\
    T_{\rm R} & = \frac{2\pi}{\alpha r^{\beta-1}}\,.
\end{align}
which demonstrates that the Rortex correctly measures the rotational part of this flow, while the other two criteria are both biased by the presence of intrinsic shears. 

It is also interesting to notice that the Rortex criterion can in principle also measure the rotational part of an irrotational vortex (i.e. when $\beta = -1$). It is well know that both the vorticity and the swirling strength values are always zero in the presence of an irrotational vortex, as also shown in Eq.\,(\ref{eq:criteria_power_vortex}). The vorticity is null because of the $1+\beta$ term, while the $\sqrt{\beta}$ term in the swirling strength would render the eigenvalue of the velocity gradient tensor purely real, hence $\lambda = 0$. The Rortex criterion does not suffer from these problems. However, in practical applications, we use Eq.\,(\ref{eq:Rortex}) to numerically compute its value. Therefore, if $\omega=0$ and $\lambda = 0$, then also $R=0$.

Finally, we test the mathematical criteria on a Lamb-Oseen vortex defined through Eq.\,(\ref{eq:lamb_oseen}). Without loss of generality, we set $\alpha = 1.0$, $r_{\rm max} = 1.0$, and $v_{\rm max} = 2/3$, so that the velocity field in Cartesian coordinates reads,
\begin{align}
    v_x & = - \frac{y}{r^2}\left(1-\exp{(-r^2)}\right)\,, \nonumber\\
    v_y & = + \frac{x}{r^2}\left(1-\exp{(-r^2)}\right)\,, \label{eq:lamb_oseen_simple}    
\end{align}
and the analytical rotational period is,
\begin{equation}
    T = \frac{2\pi r^2}{1-\exp{(-r^2)}}\,.\label{eq:lamb_oseen_simple_period}
\end{equation}
The calculations needed for the vorticity, swirling strength, and Rortex are slightly more involved, finally result in,
\begin{align}
    & \omega = 2 \exp{(-r^2)}\,, \nonumber \\
    & \lambda = \frac{2\exp{(-r^2)}}{r^2}\sqrt{\Big(\exp{(r^2)}- 1\Big)\Big(1 + 2r^2 - \exp{(r^2)}\Big)}\,, \nonumber \\
    & R = \frac{2}{r^2}\Big(1 - \exp{(-r^2)}\Big)\,.
\end{align}
Similar to the generalized rotational vortex model, each criterion yields a different value and therefore also a different period of rotation. One can check with Eq.\,(\ref{eq:period_criteria}) that the only quantity predicting the correct value for the rotation period is the Rortex criterion. Therefore the Rortex is the only reliably criterion for the extraction of physical information on the Lamb-Oseen vortex from the velocity field.


\section{Algorithm pseudo-code}
\label{app:algorithm}

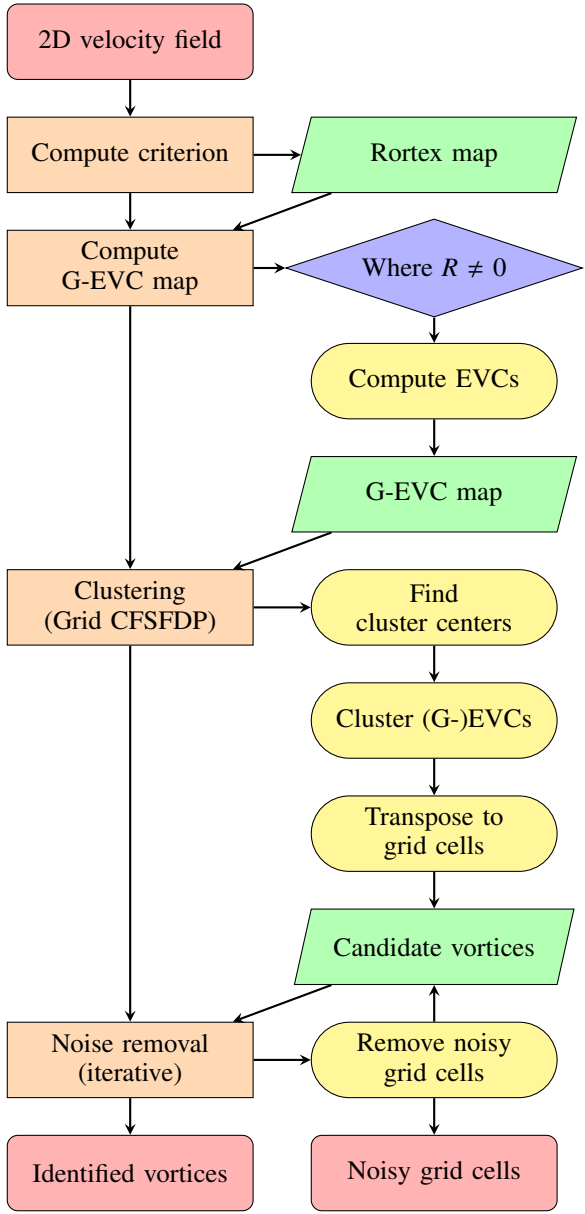
\begin{figure}
\centering
\begin{tikzpicture}[node distance=1.5cm]

\node (input) [io] {2D velocity field};
\node (step1) [function, below of=input] {Compute criterion};
\node (step1a) [result, right of=step1, xshift=2.5cm] {Rortex map};
\node (step2) [function, below of=step1] {Compute \\ G-EVC map};
\node (step2a) [condition, right of=step2, xshift=2.5cm] {Where $R\neq0$};
\node (step2b) [subfunction, below of=step2a] {Compute EVCs};
\node (step2c) [result, below of=step2b] {G-EVC map};
\node (step3) [function, below of=step2, yshift=-3cm] {Clustering \\ (Grid CFSFDP)};
\node (step3a) [subfunction, right of=step3, xshift=2.5cm] {Find\\cluster centers};
\node (step3b) [subfunction, below of=step3a] {Cluster (G-)EVCs};
\node (step3c) [subfunction, below of=step3b] {Transpose to\\ grid cells};
\node (step3d) [result, below of=step3c] {Candidate vortices};
\node (step4) [function, below of=step3, yshift=-4.5cm] {Noise removal \\ (iterative)};
\node (step4a) [subfunction, right of=step4, xshift=2.5cm] {Remove noisy \\ grid cells};
\node (output) [io, below of=step4] {Identified vortices};
\node (output2) [io, below of=step4a] {Noisy grid cells};

\draw [arrow] (input) -- (step1);
\draw [arrow] (step1) -- (step2);
\draw [arrow] (step2) -- (step3);
\draw [arrow] (step3) -- (step4);
\draw [arrow] (step4) -- (output);
\draw [arrow] (step4a) -- (output2);

\draw [arrow] (step1) -- (step1a);
\draw [arrow] (step1a) -- (step2);

\draw [arrow] (step2) -- (step2a);
\draw [arrow] (step2a) -- (step2b);
\draw [arrow] (step2b) -- (step2c);
\draw [arrow] (step2c) -- (step3);

\draw [arrow] (step3) -- (step3a);
\draw [arrow] (step3a) -- (step3b);
\draw [arrow] (step3b) -- (step3c);
\draw [arrow] (step3c) -- (step3d);
\draw [arrow] (step3d) -- (step4);

\draw [arrow] (step4) -- (step4a);
\draw [arrow] (step4a) -- (step3d);

\end{tikzpicture}
\caption{Flowchart algorithm of the SWIRL code.}
\label{fig:algorithm}
\end{figure}
The algorithm presented in Sect.\,\ref{sect:method} is shown in a diagrammatic fashion in Fig.\,\ref{fig:algorithm}. The identification process is performed over a two-dimensional velocity field, which is the input of the algorithm. We also provide an animation of the step-by-step process on a low resolution test case\footnote{The animation is available at \url{https://www.aanda.org}.
}.

The first goal is to compute the mathematical criterion that is used in the method, which is the Rortex. The Rortex can be computed on the entire grid of the two-dimensional velocity field\footnote{Except on the boundary layers if using a centered, second order finite difference derivatives to build the velocity gradient tensor.} according to Eq.\,(\ref{eq:Rortex}) for different parameters and with different stencils of grid cells. This step corresponds to the top panel of Fig.\,\ref{fig:demo_doublevortex}.

Once the Rortex map is obtained, one proceeds with the computation of the (G-)EVC map. For that, one selects the grid cells where the velocity field is characterized by some degree of curvature, that is $R\neq0$, and computes the associated estimated radial direction (Eq.\,\ref{eq:estimated_direction}) and curvature radius (Eq.\,\ref{eq:estimated_radius}). The EVCs are computed by combining these two estimated quantities, and the collection of EVC points forms the EVC map. The G-EVC map can then be obtained following the procedure explained in Sect.\,\ref{subsubsec:grid_adaptation}. The underlying idea is that the G-EVC map should show large positive (negative) values close to the center of counter-clockwise (clockwise) vortices, as shown in the bottom panel of Fig.\,\ref{fig:demo_doublevortex}.

At this point, one can employ the grid version of the CFSFDP algorithm to automatically cluster EVC points and find candidate vortices. The first step is to characterize each G-EVC by its local density $\rho$ and spacing $\delta$ defined in Eqs.\,(\ref{eq:density_grid}) and (\ref{eq:distance}), respectively. Then, one finds the cluster centers using the thresholds for $\delta$, $\rho$, or $\gamma = \rho \delta$ given as parameters. These parameters must be tweaked according to the characteristics of the flow for optimal results. An example of this selection process is given in Fig.\,\ref{fig:demo_doublevortex_decision}. Once the cluster centers have been identified, the clustering of the remaining EVCs follows the process described in Sect.\,\ref{subsubsec:vortex_identification}. At the end, every cluster of EVCs corresponds to a groups of grid cells characterized by $R \neq 0$ which form the different candidate vortices. Figure \ref{fig:demo_doublevortex_nonoise} shows two candidate vortices obtained from the two main clusters of (G-)EVCs shown in the bottom panel of  Fig.\,\ref{fig:demo_doublevortex}.

Given a set of candidate vortices, the last step consists in an iterative noise removal procedure. In this way, one can discard misidentified candidate vortices or remove noisy grid cells that were wrongly associated to a vortex. The iterative procedure is described in Sect.\,\ref{subsubsec:noise_removal}. The final outputs of the algorithm are the identified vortices and the grid cells classified as noise, as shown for example in Fig.\,\ref{fig:demo_doublevortex_result}. Each vortex consist in a set of grid cells which EVCs are sufficiently close together, meaning that the input velocity field defined on these grid cells is performing a global rotation around a common axis.   
\end{appendix}

%
%

\end{document}